\documentclass[12pt,english]{article}
\usepackage[latin1]{inputenc}
\usepackage{geometry}
\usepackage{amssymb,amsmath}

\usepackage{babel}
\usepackage{graphics}

\renewcommand{\thefootnote}{\fnsymbol{footnote}}
\newlength{\pubnumber} 

\catcode`\@=11
\@addtoreset{equation}{section}

\def\section{\@startsection{section}{1}{\z@}{3.5ex plus 1ex minus .2ex}
 {2.3ex plus .2ex}{\large\bf}}
\def\subsection{\@startsection{subsection}{2}{\z@}{2.3ex plus .2ex}
 {2.3ex plus .2ex}{\bf}}

\usepackage{amssymb}
\usepackage{epsfig}
\usepackage{cite}
\newcommand{\ba}{\begin{eqnarray}}
\newcommand{\ea}{\end{eqnarray}}
\newsavebox{\Eachterw}
\savebox{\Eachterw}(7,1)[bl]
{\put(0,0){\line(1,0){7}}
 \put(2,0){\line(0,1){1}}
\multiput(0,0)(1,0){8}{\circle*{0.2}}
\put(2,1){\circle*{0.2}}
\put(-0.125,-0.5){$\alpha_{1}$}
\put(0.875,-0.5){$\alpha_{2}$}
\put(1.875,-0.5){$\alpha_{3}$}
\put(2.875,-0.5){$\alpha_{4}$}
\put(3.875,-0.5){$\alpha_{5}$}
\put(4.875,-0.5){$\alpha_{6}$}
\put(5.875,-0.5){$\alpha_{7}$}
\put(6.875,-0.5){$\alpha_{0}$}
\put(1.875,1.5){$\alpha_{8}$}
}

\begin{document}
\begin{titlepage}
\samepage{
\setcounter{page}{1}
\rightline{CERN--PH--TH/2011--10}
\rightline{LPTENS--11/02}
\rightline{LTH--897}
\vfill
\begin{center}
 {\Large \bf Conformal Aspects of Spinor-Vector Duality}
\vfill
 {\large  Alon E. Faraggi$^1$, 
          Ioannis Florakis$^{2,3}$ ,
			Thomas Mohaupt$^1$ \\
and\\
	  Mirian Tsulaia$^1$\footnote{Associate member of the Centre for Particle Physics and Cosmology, Ilia State University, 0162 Tbilisi, Georgia}
\\
\vspace{.3in}
}

{$^1$} { Department of Mathematical Sciences,
		University of Liverpool,     \\
                Liverpool L69 7ZL, United Kingdom}\\ \vspace{2mm}
          {\small \emph{faraggi@amtp.liv.ac.uk~,~Thomas.Mohaupt@liv.ac.uk~,~tsulaia@liv.ac.uk} }\\ \vspace{4mm}
                
{$^2$} { Theory Division - CERN, \\ CH-1211 Geneva 23, Switzerland\\}  \vspace{2mm} {\small \emph{Ioannis.Florakis@cern.ch}}\\               
                
\vspace{4mm}
{ $^{3}$ Laboratoire de Physique Th\'eorique,
Ecole Normale Sup\'erieure, \\ 24 rue Lhomond, F-75231 Paris cedex 05, France\\}

\end{center}
\vfill
\begin{abstract}
  {\rm 
			We present a detailed study of various aspects of Spinor-Vector duality in Heterotic
string compactifications and expose its origin in terms of the internal conformal field theory. In particular, we illustrate the main features of the duality map by using simple toroidal orbifolds preserving $\mathcal{N}_4=1$ and $\mathcal{N}_4=2$ spacetime 
supersymmetries in four dimensions. We explain 
how the duality map arises in this context by turning on 
special values of the Wilson lines around the compact cycles of the manifold. 
 We argue that in models with $\mathcal{N}_4=2$ spacetime supersymmetry, 
the interpolation between the Spinor-Vector dual vacua 
can be  continuously realized.  We trace the origin of 
the Spinor-Vector duality map to the presence of underlying $N=(2,2)$ and $N=(4,4)$ SCFTs, 
and explicitly show that the induced spectral-flow in the twisted sectors is responsible for the observed duality. 
The isomorphism between current algebra representations gives 
rise to a number of chiral character identities, reminiscent of the recently-discovered MSDS symmetry.
}
\end{abstract}
\smallskip}
\end{titlepage}

\renewcommand{\thefootnote}{\arabic{footnote}}
\setcounter{footnote}{0}

\def\l{\label}
\def\beq{\begin{equation}}
\def\eeq{\end{equation}}
\def\beqn{\begin{eqnarray}}
\def\eeqn{\end{eqnarray}}

\def\ie{{\it i.e.}}
\def\eg{{\it e.g.}}
\def\half{{\textstyle{1\over 2}}}
\def\third{{\textstyle {1\over3}}}
\def\quarter{{\textstyle {1\over4}}}
\def\m{{\tt -}}
\def\p{{\tt +}}

\def\slash#1{#1\hskip-6pt/\hskip6pt}
\def\slk{\slash{k}}
\def\GeV{\,{\rm GeV}}
\def\TeV{\,{\rm TeV}}
\def\y{\,{\rm y}}
\def\SM{Standard-Model }
\def\SUSY{supersymmetry }
\def\SSSM{supersymmetric standard model}
\def\vev#1{\left\langle #1\right\rangle}
\def\l{\langle}
\def\r{\rangle}

\def\Htw{{\tilde H}}
\def\chibar{{\overline{\chi}}}
\def\qbar{{\overline{q}}}
\def\ibar{{\overline{\imath}}}
\def\jbar{{\overline{\jmath}}}
\def\Hbar{{\overline{H}}}
\def\Qbar{{\overline{Q}}}
\def\abar{{\overline{a}}}
\def\alphabar{{\overline{\alpha}}}
\def\betabar{{\overline{\beta}}}
\def\tautwo{{ \tau_2 }}
\def\thetatwo{{ \vartheta_2 }}
\def\thetathree{{ \vartheta_3 }}
\def\thetafour{{ \vartheta_4 }}
\def\ttwo{{\vartheta_2}}
\def\tthree{{\vartheta_3}}
\def\tfour{{\vartheta_4}}
\def\ti{{\vartheta_i}}
\def\tj{{\vartheta_j}}
\def\tk{{\vartheta_k}}
\def\calF{{\cal F}}
\def\smallmatrix#1#2#3#4{{ {{#1}~{#2}\choose{#3}~{#4}} }}
\def\ab{{\alpha\beta}}
\def\Minv{{ (M^{-1}_\ab)_{ij} }}
\def\bone{{\bf 1}}
\def\ii{{(i)}}
\def\V{{\bf V}}
\def\b{{\bf b}}
\def\N{{\bf N}}
\def\t#1#2{{ \Theta\left\lbrack \matrix{ {#1}\cr {#2}\cr }\right\rbrack }}
\def\C#1#2{{ C\left\lbrack \matrix{ {#1}\cr {#2}\cr }\right\rbrack }}
\def\tp#1#2{{ \Theta'\left\lbrack \matrix{ {#1}\cr {#2}\cr }\right\rbrack }}
\def\tpp#1#2{{ \Theta''\left\lbrack \matrix{ {#1}\cr {#2}\cr }\right\rbrack }}
\def\l{\langle}
\def\r{\rangle}

\def\La{\Lambda}
\def\te{\theta}


\def\inbar{\,\vrule height1.5ex width.4pt depth0pt}

\def\IC{\relax\hbox{$\inbar\kern-.3em{\rm C}$}}
\def\IQ{\relax\hbox{$\inbar\kern-.3em{\rm Q}$}}
\def\IR{\relax{\rm I\kern-.18em R}}
 \font\cmss=cmss10 \font\cmsss=cmss10 at 7pt
\def\IZ{\relax\ifmmode\mathchoice
 {\hbox{\cmss Z\kern-.4em Z}}{\hbox{\cmss Z\kern-.4em Z}}
 {\lower.9pt\hbox{\cmsss Z\kern-.4em Z}}
 {\lower1.2pt\hbox{\cmsss Z\kern-.4em Z}}\else{\cmss Z\kern-.4em Z}\fi}

\def\AEF{A.E. Faraggi}
\def\NPB#1#2#3{{Nucl.\ Phys.}\/ B {\bf #1}, #3 (#2)}
\def\PLB#1#2#3{{Phys.\ Lett.}\/ B {\bf #1}, #3 (#2)}
\def\PRD#1#2#3{{Phys.\ Rev.}\/ D {\bf #1}, #3 (#2)}
\def\PRL#1#2#3{{Phys.\ Rev.\ Lett.}\/ {\bf #1}, #3 (#2)}
\def\PRP#1#2#3{{Phys.\ Rep.}\/ {\bf#1}, #3 (#2)}
\def\MODA#1#2#3{{Mod.\ Phys.\ Lett.}\/ A {\bf #1}, #3 (#2)}
\def\IJMP#1#2#3{{Int.\ J.\ Mod.\ Phys.}\/ A {\bf #1}, #3 (#2)}
\def\nuvc#1#2#3{{Nuovo Cimento}\/ {\bf #1A}, #3 (#2)}
\def\JHEP#1#2#3{{JHEP} {\bf #1}, #3 (#2)}
\def\EJP#1#2#3{{ Eur.\ Phys.\ Jour.}\/ C {\bf #1}, #3 (#2)}
\def\MPLA#1#2#3{{ Mod.\ Phys.\ Lett.}\/ A {\bf #1}, #3 (#2)}
\def\IJMPA#1#2#3{{ Int.\ J.\ Mod.\ Phys.}\/ A {\bf #1}, #3 (#2)}

\def\etal{{\it et al\/}}

\hyphenation{su-per-sym-met-ric non-su-per-sym-met-ric}
\hyphenation{space-time-super-sym-met-ric}
\hyphenation{mod-u-lar mod-u-lar-in-var-i-ant}


\setcounter{footnote}{0}
\section{Introduction}
\bigskip

String theory provides a detailed framework to explore the unification of the gauge and gravitational interactions. Progress in this endeavour mandates 
both a deeper understanding of the various mathematical structures underlying 
the theory as well as the development of phenomenological models 
that aspire to make contact with observation. It is clear that a deeper understanding of the structure and various dualities underlying such models may further elucidate their basic properties. 

Over the last few years a novel `Spinor-Vector' duality map has been observed in 
the massless spectra of Heterotic $\mathbb{Z}_2\times \mathbb{Z}_2$ compactifications under the 
exchange of the vectorial and spinorial representations of the $SO(10)$ GUT gauge group. 
The initial observation of Spinor-Vector duality  \cite{fkr,fkr2,fkr3} in $\mathbb{Z}_2\times \mathbb{Z}_2$ symmetric 
orbifold vacua was made by using the powerful and systematic \cite{gkr,fknr} classification methods of the free fermionic 
formulation, by means of numerical and analytical techniques. 
A special property of $\mathbb{Z}_2$-type orbifolds is that $\mathcal{N}_4=1$ twisted sectors 
inherit the structure of the $\mathcal{N}_4=2$ ones. In particular, this implies that, as the 
duality map is realised internally in each twisted sector, it can be seen to 
hold in $T^4/\mathbb{Z}_2$ vacua \cite{fkr3} as well. Analytic proof of the Spinor-Vector 
duality within the framework of the fermionic construction was given in refs. \cite{fkr2,fkr3, cfkr} for the $\mathbb{Z}_2$- as well as the $\mathbb{Z}_2\times \mathbb{Z}_2$- case.

In ref. \cite{aft} the $E_8\times E_8$ heterotic string compactified on a symmetric, non-freely-acting $T^2\times(T^4/ \mathbb{Z}_2)$ orbifold was considered. 
This was then followed by two additional freely-acting $\mathbb{Z}_2'\times\mathbb{Z}_2''$ orbifolds \cite{Faraggi:2002qh}, each correlating 
the charges of an $E_8$-factor to a half-shift along a compact cycle of the untwisted $T^2$. The resulting vacuum was characterized by $\mathcal{N}_4=2$ spacetime supersymmetry and 
its genus-1 partition function included 8 independent orbits and, therefore, 7 discrete torsions. Within that framework, Spinor-Vector duality
 was seen to arise from different choices for the values of these discrete torsions.


At a deeper level, Spinor-Vector duality in $\mathcal{N}_4=1$ theories can be seen to be a remnant of the
spontaneous breaking of $N=(2,2)$ worldsheet supersymmetry to $N=(0,2)$. The $N=(2,2)$-constructions \cite{Gepner}, 
correspond to compactifications on Calabi-Yau surfaces that extend the gauge symmetry 
from $SO(10)\times U(1)$ to $E_6$. The preservation of the global 
left-moving $N=2$ superconformal algebra in this setting, corresponds to the self-dual 
case under the duality map, in the sense that both vectorial and spinorial 
representations of the $SO(10)\subset E_6$ are then massless. This reflects the fact that the matter 
representations of  $SO(10)$, namely the $\textbf{16}$ (spinorial) and $\textbf{10}$ (vectorial), 
together with the singlet $\textbf{1}$, do fit nicely into the $\textbf{27}$ representation of the enhanced symmetry group $E_6$.
Then one may give non-vanishing mass to either the vectorial or the spinorial 
representations (or both) by turning on suitable discrete Wilson lines. The resulting vacua will be dual to each other through the above Spinor-Vector duality map.


In the case of $T^4/\mathbb{Z}_2$ Heterotic compactifications with $\mathcal{N}_4=2$ supersymmetry, 
the  $\hat{c}=6$ internal CFT breaks into a $\hat{c}=2$, $N=2$ superconformal system in terms of 
2 free compact (super-)coordinates while the 4 remaining internal coordinates form an $\hat{c}=4$, $N=4$ system \cite{BanksDixon}. At the enhanced symmetry point, where $SO(12)\times SU(2)\rightarrow E_7$, the 
global left-moving internal CFT becomes enhanced into $N=(4,4)$. The result of this 
enhancement is twofold. First of all it guarantees the presence of $SO(12)$ spinorials 
and vectorials (always accompanied by singlets) in the massless spectrum, marking this as 
the self-dual point under the duality map. Secondly, it ensures the existence of an $N=4$ spectral-flow 
operator, transforming the (spinorial) $\textbf{32}$ representation of $SO(12)$ into the 
vectorial $\textbf{12}$ plus the singlet $\textbf{1}$. This spectral-flow is 
responsible for the fact that the number of massless degrees of freedom in the spinorial 
representation is the same as that in the vectorial and singlets. As before, by turning 
on suitable Wilson lines, either the spinorial or the vectorial (plus singlet) 
representations of $SO(12)$ may acquire non-vanishing mass, which manifests itself as the observed Spinor-Vector 
duality. However, the matching of the number of massless degrees of freedom between these 
representations at the points of symmetry enhancement, ensures that these numbers will continue to be equal as the theory is deformed away from these critical points.


The initial observation and study of the Spinor-Vector duality map in \cite{fkr,fkr2,fkr3,cfkr}
was made within the framework of the fermionic construction \cite{fff}. 
However, even though such formulations at the fermionic point are very effective for scanning the space of
phenomenologically attractive vacua \cite{ffmodels,ffmodels2}, they are typically  
limited only to particular points in moduli space, where the compactification radii and other 
background fields take specific values. This limited description may sometimes obscure the underlying 
physics and may mask the true CFT structure and origin of various maps, such as the Spinor-Vector duality map. 
For this purpose, it is important to deform these theories away from the `special' fermionic points, or 
to directly develop constructions where the duality is manifested at generic points in moduli space. 
This will be achieved partially in the present paper, where our arguments will be valid at a generic point in moduli space.


The purpose of this paper is to further investigate the CFT nature of Spinor-Vector duality and explicitly demonstrate how 
the duality results from the spectral flow of global $N=2$ or $N=4$ SCFTs, that arise from the embedding of the spin connection of Type II theories into the gauge connection of Heterotic ones.

An interesting discovery is that the relevant spectral-flow operator in the twisted 
sector is identical to the operator generating the recently 
discovered \emph{Massive Spectral boson-fermion Degeneracy Symmetry} (MSDS) \cite{MSDS}, \cite{ReducedMSDS}. The MSDS structure typically arises chirally in the worldsheet 
supersymmetric sector of exotic 2d string constructions living at special extended symmetry 
points in the moduli space. In those constructions, it stems from a special breaking of the global $N=2$ SCFT generating spacetime supersymmetry, to a novel enhanced current algebra, thus, implying a very 
specific (spontaneous) breaking of spacetime supersymmetry. The trademark of 
these constructions is that all massive bosonic and fermionic modes are matched, similarly to the case of conventional supersymmetry. However, massless bosonic and fermionic modes remain unpaired:
\begin{align}
	n_b-n_f ~ \left\{
\begin{array}{l l}
	=0 & \textrm{for}~m>0 \\
	\neq 0 & \textrm{for}~m=0
\end{array}\right. .
\end{align}

The paper is organized as follows:
\\
In Section \ref{SVreview} we present an overview of Spinor-Vector duality in terms of a 
specific $\mathcal{N}_4=2$ model in which the duality is exhibited in a clear and simple way. 
In particular, we start with an $\mathcal{N}_4=2$ Heterotic compactification 
on $S^1\times \tilde{S}^1\times(T^4/\mathbb{Z}_2)$ and then consider an additional freely-acting $\mathbb{Z}_2'$-orbifold, 
correlating the Cartan charges of the full $E_8\times E_8$ with a half-shift along the compact $S^1$-circle. We demonstrate how the duality map arises within this description for different choices of the discrete torsion. 

In Section \ref{WilsonLines}, we proceed to demonstrate how the freely-acting $\mathbb{Z}_2'$ can be 
reformulated as a Wilson line background around the $S^1$-circle, where now the 
choice of discrete torsion is translated into the specific choices for the value of the Wilson line. 
 We show that within the moduli space of $\mathcal{N}_4=2$ vacua, the interpolations between vacua with massless $SO(12)$-spinorials and 
those with massless vectorials (plus singlets) can be continuously performed. 
This situation differs substantially from the $\mathcal{N}_4=1$ 
compactifications on $T^6/(\mathbb{Z}_2\times\mathbb{Z}_2)$, where the 
analogous Wilson lines do not correspond to invariant marginal operators that may be used to perturb the $\sigma$-model and, hence, can only take specific discrete values. 

In Section \ref{SpFlow}, we analyze the superconformal properties of $N=(2,2)$ and $N=(4,4)$ internal CFTs and 
illuminate the true source and structure of the Spinor-Vector duality map. 
In particular, we show how embedding the $N=2$ and $N=4$ SCFTs of Type II theories into the left-moving 
bosonic side of Heterotic string theory, gives rise to a spectral-flow which is 
responsible for transforming the spinorial representations of $SO(10)$ and $SO(12)$ into the 
vectorial and singlet representations. We explicitly construct the spectral-flow operator in each 
case and we demonstrate the induced isomorphism between the representations of the current algebra. In particular, this explains why the number of massless degrees of freedom remains unchanged under the duality map.

In Section \ref{NarainLattices}, we give a complementary discussion from the Hamiltonian 
perspective, using the underlying Narain lattice. In this formulation the interpolation between models, 
and the patterns of symmetry breaking and symmetry enhancement is particularly transparent. In Section \ref{K3section} we  briefly 
review how the orbifold models considered in this paper are related to generic K3 compactifications of the $E_8 \times E_8$ heterotic string.
 
Finally, in Section \ref{Conclusions} we present our conclusions and directions for future research.


\section{Review of Spinor-Vector Duality}\label{SVreview}

In this Section we provide an overview of Spinor-Vector duality. After introducing our convensions 
we directly proceed with the definition of a very specific model, which will serve as a working example in which the structure of the duality map will be illustrated.

\subsection{Generalities and Conventions}

Throughout the paper we set $\alpha'=2$, in order to avoid additional $\sqrt{2}$-factors in the 
exponents of vertex operators. In particular, this implies that a free complex fermion $\Psi(z)$ is bosonized in terms of a real compact boson $\Phi(z)$ as:
\begin{align}
	\left. \begin{array}{l}
				\Psi(z) = e^{i\Phi}\\
				\Psi^\dagger(z)= e^{-i\Phi}\\
				\end{array}\right\} ~\leftrightarrow~\Psi\Psi^\dagger(z) = i\partial\Phi(z).
\end{align}  
In these convensions, the above equivalence can be realized at a bosonic radius $R=1$, commonly refered to as the `fermionic point'. More generally, a vertex operator $e^{iq\Phi}$ has conformal weight $\Delta=q^2/2$.

For Heterotic theories, we adopt the usual convention in which spacetime fermions arise from spin 
fields of the right-moving (anti-holomorphic) sector. Hence, the right-movering 
sector is characterized by a local $N=1$ superconformal algebra, 
which results from gauge fixing the (super-)reparametrization invariance, whereas the left-moving (holomorphic) sector is similar to the bosonic string and contains the gauge degrees of freedom.

In the bosonic formulation \cite{heterotic} of the $E_8\times E_8$ Heterotic 
string\footnote{For some recent developments on the phenomenology of Heterotic orbifolds see, 
for example, 
refs \cite{Extra}, \cite{recentheterotic} and references therein.}, the conformal anomaly in the left-moving sector is canceled by introducing $16$ additional bosons compactified on the $E_8\times E_8$ chiral root lattice.

Here we will rather use the fermionic formulation of the Heterotic string, where the left-moving 
conformal anomaly is instead canceled by the insertion of 16 
free (complex) worldsheet fermions $\Psi^{A},\lambda^A$, with $A=1,\ldots, 8$. 
If all $16$ complex fermions are assigned the same (real) boundary conditions, the sum over the spin structures yields the $Spin(32)/\mathbb{Z}_2$ lattice:
\begin{align}
	\Gamma_{16}= \frac{1}{2}\sum\limits_{\gamma,\delta=0,1}{\frac{\theta[^\gamma_\delta]^{16}}{\eta^{16}}}.
\end{align}
On the other hand, by grouping the complex fermions into two 
groups of eight, $\{\Psi^A\}$, $\{\lambda^A\}$, such that the 
fermions in each group share common boundary conditions and 
summing over the (independent) possible boundary conditions of each group, one obtains the representation of the $E_8\times E_8$ lattice in terms of Jacobi $\theta$-functions:
\begin{align}
	\Gamma_{E_8\times E_8}(\tau) = \left[\frac{1}{2}\sum\limits_{k,\ell=0,1}{\frac{\theta[^k_\ell]^{8}}{\eta^8}}\right]~\left[\frac{1}{2}\sum\limits_{\rho,\sigma=0,1}{\frac{\theta[^\rho_\sigma]^{8}}{\eta^8}}\right].
\end{align}
The modular invariant partition function of the ten-dimensional $E_8\times E_8$ Heterotic string then becomes:
\begin{align}
	Z_{E_8\times E_8} = \frac{1}{\tau_2^{4}\eta^8\bar\eta^{8}}~\left[\frac{1}{2}\sum\limits_{\bar{a},\bar{b}=0,1}{(-)^{\bar{a}+\bar{b}+\bar{a}\bar{b}}~\frac{\bar\theta[^{\bar{a}}_{\bar{b}}]^4}{\bar\eta^4}}\right]~ \Gamma_{E_8\times E_8}(\tau).
\end{align}
It is convenient to decompose the spectrum into characters of the global the $SO(2n)$ worldsheet current algebra realized in terms of worldsheet fermions:
\begin{align}
	Z_{E_8\times E_8} = \frac{1}{\tau_2^{4}\eta^8\bar\eta^{8}}~\left(\bar{V}_8-\bar{S}_8\right)~ \left(O_{16}+S_{16}\right)\left(O_{16}+S_{16}\right),
\end{align}
where:
\begin{align}\nonumber
	O_{2n} &= \frac{1}{2}\left(~\frac{\theta_3^n}{\eta^n} + \frac{\theta_4^n}{\eta^n}~\right), \\ \nonumber
	V_{2n} &= \frac{1}{2}\left(~\frac{\theta_3^n}{\eta^n} - \frac{\theta_4^n}{\eta^n}~\right), \\ \nonumber
	S_{2n} &= \frac{1}{2}\left(~\frac{\theta_2^n}{\eta^n} + e^{-i\pi n/2} \frac{\theta_1^n}{\eta^n}~\right), \\
	C_{2n} &= \frac{1}{2}\left(~\frac{\theta_2^n}{\eta^n} - e^{-i\pi n/2} \frac{\theta_1^n}{\eta^n}~\right),
\end{align}
and $\theta_1\equiv\theta[^1_1]$, $\theta_2\equiv\theta[^1_0]$, $\theta_3\equiv\theta[^0_0]$, $\theta_4\equiv\theta[^0_1]$.

Finally, Appendix \ref{PartitionFunction} contains a detailed calculation of the partition 
function for the simple $\mathcal{N}_4=2$ model, that is used to illustrate the Spinor-Vector duality map. Moreover, in Appendix \ref{OPEs}, we summarize useful OPEs involving spin-fields of the $SO(N)$ current algebra.


\subsection{Spinor-Vector duality in a $T^4/\mathbb{Z}_2$ orbifold}\label{ToyModelReview}

We are now ready to present the structure of Spinor-Vector duality in terms of an explicit example, 
realized in a simple compactification of the $E_8\times E_8$ heterotic 
string on a $T^2\times T^4/\mathbb{Z}_2$ orbifold. Because 
twisted sectors of $\mathcal{N}_4=1$, $T^6/\mathbb{Z}_2\times\mathbb{Z}_2$ constructions 
inherit the $\mathcal{N}_4=2$ structure of twisted sectors in $T^4/\mathbb{Z}_2$ vacua, 
it will be sufficient for our purposes to analyze the latter case in some detail. For simplicity, 
we will consider the case where the $T^2$-torus factorizes\footnote{It should be noted 
that this factorizability of $T^2$ is not necessary for the general results of 
this section. It can be shown that Spinor-Vector duality continues to persist 
for generic values of the $T^2$-moduli.} into two independent circles $S^1(R)\times \tilde{S}^1(\tilde{R})$, parametrized by the $X^8$ and $X^9$ internal coordinates, respectively. 

The orbifold group acts on the internal coordinates of $T^4$ and on their fermionic superpartners as:
\begin{align} \label{Z2twist}
	g: \left\{\begin{array}{c}
	\bar\psi^I(\bar{z}) \rightarrow -\bar\psi^I(\bar{z}) \\
	X^I(z,\bar{z}) \rightarrow -X^I(z,\bar{z}) \\
\end{array}\right. ~~,~~\textrm{for}~I=4,5,6,7 
\end{align}
whereas the remaining two internal coordinates $X^8, X^9$, parametrizing $S^1\times\tilde{S}^1$, are invariant.

The standard embedding of the point group in the gauge sector is realized as a twist in the boundary conditions of two complex fermions associated to the gauge degrees of freedom:
\begin{align} \label{StdEmb}
	g: \Psi^A \rightarrow -\Psi^A ~~,~~\textrm{for}~A=1,2,
\end{align}
while the remaining $14$ left-moving fermions remain untwisted. Note that in the bosonic 
formulation, where the complex fermions $\Psi$ are bosonized as $\Psi(z)=e^{iH(z)}$, the orbifold action becomes realized as a half-shift in the compact bosonic coordinate $H(z)\rightarrow H(z)+\pi$.

It is convenient to decompose the $SO(2n)$ characters into characters of lower-dimensional subgroups, in which the $\mathbb{Z}_2$-orbifold action is diagonal:
\begin{align}
	\bar{V}_8-\bar{S}_8 = \bar{V}_4 \bar{O}_4 + \bar{O}_4 \bar{V}_4 - \bar{S}_4\bar{S}_4 - \bar{C}_4\bar{C}_4 
\end{align}
This is the standard decomposition of the $SO(8)$-little group into representations of $SO(4)\times SO(4)$, 
where the first $SO(4)$-factor corresponds to the 2 transverse worldsheet fermions $\psi^\mu$, with $\mu=2,3$ and the 
fermionic superpartners of the two untwisted toroidal coordinates $X^{8,9}$ 
parametrizing $S^1\times \tilde{S}^1$. The second $SO(4)$-factor will correspond to the supercoordinates $( X^I, \psi^I,~I=4,5,6,7)$, that are twisted under the $\mathbb{Z}_2$-action.

Similarly, the global $SO(16)$-characters in the left-moving sector can be decomposed into characters of $SO(4)\times SO(12)$:
\begin{align}
	O_{16}+S_{16} = O_{4}O_{12} + V_{4}V_{12} + S_{4}S_{12}+C_{4}C_{12}~,
\end{align}
where the $SO(4)$-subgroup is realized by the 2 complex fermions $\Psi^{1,2}$ that transform under the 
orbifold action, while the $SO(12)$-subgroup is associated to the 
remaining $6$ fermions $\Psi^{3,\ldots,8}$ on which the orbifold embedding is trivial. 
Since the orbifold action does not twist the remaining fermions $\lambda^A$, associated to $E_8'$, the relevant contribution will still be in terms of $SO(16)$-characters. 

The generic action of $\mathbb{Z}_2$ on the $SO(2n)$-characters associated to the twisted fermions:
\begin{align}\nonumber
	&O_{2n} \rightarrow +O_{2n},\\ \nonumber
	&V_{2n} \rightarrow -V_{2n},\\ \nonumber
	&S_{2n} \rightarrow +e^{-i\pi n/2} S_{2n},\\
	&C_{2n} \rightarrow -e^{-i\pi n/2} C_{2n}.
\end{align}
This, of course, reflects the fact that the $O_{2n}$ representation corresponds to the vacuum 
state (and the adjoint in the first excited level), which stays invariant under a 
twist in the boundary conditions of the $SO(2n)$-fermions. On the other hand, the vectorial representation in the $SO(2n)$-current algebra is linear in the worldsheet fermions and, thus, $V_{2n}$ changes sign under the fermion twist.

This implies the following action of the $\mathbb{Z}_2$-orbifold on the right-moving $SO(4)\times SO(4)$ fermion characters:
\begin{align}
	\bar{V}_8-\bar{S}_8 \xrightarrow{~~\mathbb{Z}_2~~} \bar{V}_4 \bar{O}_4 - \bar{O}_4 \bar{V}_4 + \bar{S}_4\bar{S}_4 - \bar{C}_4\bar{C}_4 ~.
\end{align}
Already it becomes visible that the action of the $\mathbb{Z}_2$-orbifold projects out half of the gravitini, so that the compactification on  $T^4/\mathbb{Z}_2$ will describe an $\mathcal{N}_4=2$ supersymmetric vacuum.

Similarly, the orbifold action on the left-moving gauge sector transforms the $SO(4)\times SO(12)$ fermion characters of the first $E_8$ factor as:
\begin{align}
	O_{16}+S_{16} \xrightarrow{~~\mathbb{Z}_2~~} O_{4}O_{12} - V_{4}V_{12} - S_{4}S_{12}+C_{4}C_{12}~.
\end{align}

Under the action of the $\mathbb{Z}_2$-orbifold, the (untwisted) moduli space of the $\mathcal{N}_4=2$ theory is reduced down to:
\begin{align}
\label{UntwistedModuli}
		\frac{SO(16+6,6)}{SO(16+6)\times SO(6)} ~\xrightarrow{~\mathbb{Z}_2~}~ \frac{SO(4,4)}{SO(4)\times SO(4)}~\times~\frac{SO(16+2,2)}{SO(16+2)\times SO(2)}~.
\end{align}
The $\frac{SO(4,4)}{SO(4)\times SO(4)}$-factor corresponds to Lorentz boosts of the $\Gamma_{(4,4)}$ lattice associated with the $T^4$ or, equivalently, to marginal deformations with respect to the $16$ moduli $G_{IJ}, B_{IJ}$. 

Similarly, the $\frac{SO(16+2,2)}{SO(16+2)\times SO(2)}$-factor contains the 
Lorentz boosts in the $\Gamma_{(18,2)}$-lattice, which can be equivalently 
obtained from any particular point in moduli space\footnote{For example, one could define 
the theory at the so-called ``fermionic" point in moduli space, where all internal 
bosonic coordinates $X^I$ can be consistently fermionized in terms of free (complex) worldsheet fermions $i\partial X_{L,R}^I=i\bar\psi_{L,R}^I\psi_{L,R}^I$.} by 
marginally deformating with respect to the $4$ moduli in $T^2$, $G_{ij}$ and $B_{ij}$, as 
well as by turning on Wilson lines $A_{i}^a$, with $a=1,\ldots,16$ taking values along the $16$ Cartan generators of $E_8\times E_8$ and $i,j=8,9$.

Furthermore, we introduce an additional freely-acting $\mathbb{Z}_2'$-orbifold :
\begin{align}
	g'= e^{2\pi i(Q_{8}+Q_{8}')}\delta,
\end{align}
where $\delta$ is a half-shift along the $S^1(R)$ compact direction:
\begin{align}
	X^8\rightarrow X^8+\pi R,
\end{align}
and $Q_8$ and $Q_8'$ are the $U(1)$ gauge charges with respect to the generators in the Cartan 
subalgebra of $E_8$ and $E_8'$, respectively. The spinorial (or anti-spinorial) representations carry half-integer charges $Q\in \mathbb{Z}+\frac{1}{2}$, 
whereas the adjoint and vectorial representations have integer charges $Q\in \mathbb{Z}$. In the fermionic 
formulation of $E_8$ (resp. $E_8'$), the parity operator $e^{2\pi i Q_8}$ (resp. $e^{2\pi i Q_8'}$) becomes associated to 
the spin structure of the corresponding set $\{\Psi^A\}$ (resp. $\{\lambda^A\}$) of the complex worldsheet fermions, similarly to the 
spacetime fermion number $(-)^F$ for the right-moving worldsheet fermions.

The effect of the freely acting $\mathbb{Z}_2'$ is to correlate the gauge charges $Q_8$, $Q_8'$ with a half-shift along the circle $S^1(R)$ parametrized by $X^8$. 
As will be shown in the next section, this freely acting orbifold corresponds to a particular choice of the Wilson line along the $X^8$ circle and, 
as such, it can be equivalently described as a Lorentz boost of the full (untwisted) $\Gamma_{(18,2)}$-lattice. In terms of the $\Gamma_{(1,1)}$ lattice 
associated to the $X^{8}$-circle, the action of the freely-acting $\mathbb{Z}_2'$ is that of a momentum shift. Indeed, in the Hamiltonian representation, the $\Gamma_{(1,1)}(R)$-lattice takes the form:
\begin{align}
	\Gamma_{(1,1)}(R) = \sum\limits_{m,n\in\mathbb{Z}}{\Lambda_{m,n}(R)}=\frac{1}{\eta\bar\eta}\sum\limits_{m,n\in\mathbb{Z}}{q^{\frac{1}{2}P_L^2}\bar{q}^{\frac{1}{2}P_R^2} },
\end{align}
where:
\begin{align}
		P_{L,R} = \frac{m}{R}\pm \frac{nR}{2},
\end{align}
and $m,n$ are the momentum and winding numbers around the $X^8$-cicle, respectively.
Then the action of the freely-acting orbifold $\mathbb{Z}_2'$ on the $S^1$ lattice is simply a momentum shift:
\begin{align}
	\Lambda_{m,n}(R) ~\xrightarrow{~\delta~}~ (-)^m \Lambda_{m,n}(R).
\end{align}

The modular invariant partition function of the model can be decomposed into:
\begin{align}
	Z = \frac{1}{(\sqrt{\tau_2} \eta \bar\eta)^2}\left[Z_{(0,0)} + Z_{(1,0)} + Z_{(0,1)} + Z_{(1,1)}\right] ~,
\end{align}
where
\begin{align}
	Z_{(h,h')} = \frac{1}{2^2}\sum\limits_{g,g'=0,1}{Z[^{h,h'}_{g,g'}]} ~.
\end{align}
Here, $Z_{(h,h')}$ denote the sectors twisted by the $\mathbb{Z}_2\times\mathbb{Z}_2'$ orbifold and the summation over $g,g'$ incorporates the projection to invariant states. 

It is convenient to define the twisted characters:
\begin{align}
	Q_o= \overline{O}_4 \overline{V}_4 - \overline{S}_4 \overline{S}_4 ~~ &, ~~ Q_v = \overline{V}_4 \overline{O}_4 - \overline{C}_4 \overline{C}_4 \\
	P_o= \overline{O}_4 \overline{C}_4 - \overline{S}_4 \overline{O}_4 ~~ &, ~~ P_v = \overline{V}_4 \overline{S}_4 - \overline{C}_4 \overline{V}_4 ~,
\end{align}
which are the linear combinations of standard $SO(4)\times SO(4)$ characters that are eigenvectors with respect to the orbifold action.

The relative sign between the two orbits is arbitrary and is parametrized by the discrete torsion coefficient $\epsilon=\pm 1$. In terms of the discrete torsion parameter, the modular invariant partition function can be written as:
\begin{align}
	Z = \frac{1}{(\sqrt{\tau_2} \eta \bar\eta)^2} ~\frac{1}{2^2}\sum\limits_{h,g=0,1}~\sum\limits_{h',g'=0,1}{(-)^{\frac{1-\epsilon}{2}(hg'-gh')} Z[^{h,h'}_{g,g'}]},
\end{align}
where the inclusion or not of the modular invariant cocycle $(-)^{hg'-h'g}$ alternates (or not) the sign of the second orbit. 
Furthermore, in the interest of simplicity, and throughout this paper, the contribution of the spectator $\Gamma_{(1,1)}(\tilde{R})$-lattice, associated to the $X^9$-circle will be suppressed.


 The explicit calculation of the partition function is presented in considerable detail in Appendix \ref{PartitionFunction}. Here we 
will directly discuss the spectrum of this model and comment on the appearance of the Spinor-Vector duality, which will 
now become transparent. Massless states can be seen to arise from the fully untwisted sector $Z_{(0,0)}$ and from the sector $Z_{(1,0)}$ twisted under the non freely-acting $\mathbb{Z}_2$. 
On the other hand, both sectors $Z_{(0,1)}$ and $Z_{(1,1)}$, which are twisted by the freely-acting $\mathbb{Z}_2'$, are characterized by non-trivial winding, hence, rendering all states within these two sectors massive.  

The untwisted sector contains the representation
$$
	Q_v ~\Lambda_{2m,n} \Gamma^{h=0}_{(+)}~O_{12}O_4 O_{16} ,
$$
which gives rise both to the gravity multiplet as well as the space-time vector bosons generating the $SO(12)\times SO(4)\times SO(16)$ gauge symmetry. In addition, it contains
$$
	Q_o ~\Lambda_{2m,n} ~ \Gamma^{h=0}_{(+)}~V_{12}V_4 O_{16},
$$
giving rise to scalar multiplets transforming in the bi-vector representation of $SO(12)\times SO(4)$.

Let us now examine the $\mathbb{Z}_2$-twisted sector $Z_{(1,0)}$ and see how the Spinor-Vector
duality operates. We first note that massless states can only arise for vanishing momentum
and winding quantum numbers $m=n=0$ along the $S^1$-circle and only from the $P_o$-sector. 
The representations that can become massless are then :
\begin{align}
		& P_o ~\Lambda_{2m+\frac{1-\epsilon}{2},n} \left(~ \Gamma^{h=1}_{(+)} ~V_{12}C_4 O_{16}  + \Gamma^{h=1}_{(-)}~O_{12}S_4 O_{16} ~\right)\\ \nonumber
	& P_o ~\Lambda_{2m+\frac{1+\epsilon}{2},n} ~ \Gamma^{h=1}_{(+)}~S_{12}O_4 O_{16} .
\end{align}
It is then clear that, distinct choices of discrete torsion ($\epsilon=\pm1$) give mass either to the spinorial 
representation of $SO(12)$, while keeping the the vectorial and scalar representations massless, or render the spinorial massless and give mass to the vectorial and scalar representations, instead. 

Indeed, in the case with $\epsilon=+1$ the zero lattice modes
attach to the $P_o\, V_{12} C_{4}O_{16}$-representation, that produces $8$ massless $\mathcal{N}_4=2$ hypermultiplets in the 
vectorial representation $(\textbf{12},\textbf{2})$ of $SO(12)\times SO(4)$, whereas
in the case with  $\epsilon=-1$ the zero lattice modes
attach to $P_o\, S_{12} O_{4} O_{16}$,
which produces $8$ massless $N=2$ hypermultiplets in the
$(\textbf{32},\textbf{1})$ spinorial representation. Furthermore, in the case with $\epsilon=+1$ the first excited twisted lattice
modes produce $8\times 2\times 4$ massless $SO(12)$-singlets $(\textbf{1},\textbf{2})$ from the term
$P_o\, O_{12} S_{4} O_{16}$. 
A very interesting observation is that the total number of massless states (equal to $2\times 8\times 32$)
is the same in both cases $\epsilon=\pm 1$. 


\subsection{Spinor-Vector Duality map via non-trivial Wilson line backgrounds}\label{WilsonLines}

In the previous section we explicitly analysed the spectrum of a particular Heterotic model, compactified on $S^1\times \tilde{S}^1\times T^4/\mathbb{Z}_2$. There, 
an additional freely-acting $\mathbb{Z}_2'$ orbifold, correlating the gauge charges with a translation along the $S^1$-circle, was introduced and the change in the choice of the 
discrete torsion $\epsilon$, associated to the two independent modular orbits, was shown to produce the Spinor-Vector dual theory. 

However, this formulation of the duality in terms of the choice of discrete torsion is not suitable to reveal its underlying structure. 
In order to display this structure it will be convenient to rewrite the partition function in a representation where modular covariance will be manifest.
Indeed, it is straightforward to obtain the following covariant expression for the $\mathbb{Z}_2\times\mathbb{Z}_2'$ orbifold blocks:
\begin{align}\nonumber
	Z[^{h,h'}_{g,g'}]= &\left[\frac{1}{2}\sum\limits_{\bar{a},\bar{b}=0,1}{(-)^{\bar{a}+\bar{b}+\bar{a}\bar{b}}
C_{1}~\frac{\bar\theta[^{\bar{a}}_{\bar{b}}]^2\bar\theta[^{\bar{a}+h}_{\bar{b}+g}]\bar\theta[^{\bar{a}-h}_{\bar{b}-g}]}{\bar\eta^4} }\right]~\Gamma_{(4,4)}[^h_g]\\
&\times\Gamma_{(1,1)}[^{h'}_{g'}]~(-)^{h'(\ell+\sigma)+g'(k+\rho)}~\left[\frac{1}{2}\sum\limits_{k,\ell=0,1}{C_2 
\frac{\theta[^k_\ell]^{6}\theta[^{k+h}_{\ell+g}]\theta[^{k-h}_{\ell-g}]}{\eta^8}}\right]~\left[\frac{1}{2}\sum\limits_{\rho,\sigma=0,1}{\frac{\theta[^\rho_\sigma]^{8}}{\eta^8}}\right],
\end{align}
where $C_1$ and $C_2$ are modular invariant phases fixing the chiralities of the spinorial current algebra representations. 
In order to have agreement with the chirality conventions that appear in the definition of the model in the previous section, we choose:
\begin{align}\nonumber
		C_1 = (-)^{\bar{a}\bar{b}}(-)^{(\bar{a}+h)(\bar{b}+g)},\\
		C_2 = (-)^{k\ell}(-)^{(k+h)(\ell+g)}.
\end{align}
Also,
\begin{align}
		\Gamma_{(1,1)}[^{h'}_{g'}](R)= \frac{R}{\sqrt{2\tau_2}}~\sum\limits_{\tilde{m},n\in\mathbb{Z}}{e^{-\frac{\pi R^2}{2\tau_2}\left|\tilde{m}+\frac{g'}{2}+\tau\left(n+\frac{h'}{2}\right)\right|^2}} 
\end{align}
is the $(1,1)$-lattice in the Lagrangian representation with half-shifted windings. This can be easily 
verified by noting that in the $[^{h'}_{g'}]$-twisted sector the boundary conditions of the compact scalar $X^8$ can also be satisfied as:
\begin{align}\nonumber
		& X^8(\sigma^1+2\pi,\sigma^2) \sim X^8(\sigma^1,\sigma^2)+2\pi nR + h'\pi R ,\\
		& X^8(\sigma^1,\sigma^2+2\pi) \sim X^8(\sigma^1,\sigma^2)+2\pi \tilde{m}R + g'\pi R ,
\end{align} 
where $\tilde{m},n$ are the two winding numbers along the $X^8$-circle. Then, the insertion of the modular invariant 
cocycle $(-)^{h'(\ell+\sigma)+g'(k+\rho)}$ has exactly the effect of alternating the sign depending on the gauge charges $e^{2\pi i(Q_8+Q_8')}$, as is required by the freely-acting $\mathbb{Z}_2'$ action.

In what follows we will illustrate that the freely acting $\mathbb{Z}_2'$-orbifold is equivalent to a very specific choice of
 Wilson line along the $S^1$-circle. From the latter perspective, the arbitrariness in the choice of discrete torsion will be seen to correspond to a particular freedom in the choice of the Wilson lines. 

We begin by performing a double Poisson resummation to bring the $\Gamma_{(1,1)}$ lattice to its dual form
\begin{align}
		\Gamma_{(1,1)}[^{h'}_{g'}]= \frac{(1/R)}{\sqrt{2\tau_2}}~\sum\limits_{\tilde{m},n\in\mathbb{Z}}{e^{-\frac{\pi }{2\tau_2}\left(\frac{1}{R}\right)^2\left|n+\tau\tilde{m}\right|^2}(-)^{\tilde{m}g'+nh'}} ,
\end{align}
so that $h',g'$ only appear through the phase. It is now possible to completely perform the summation over $h',g'$:
\begin{align}\label{Z2orbBlocks}
	Z = \frac{1}{(\sqrt{\tau_2} \eta \bar\eta)^2}~ \frac{1}{2}\sum\limits_{h,g=0,1}{Z[^h_g]},
\end{align}
and reduce the partition function down to the sum of the orbifold blocks of the non-freely acting $\mathbb{Z}_2$:
\begin{align}\nonumber
			Z[^h_g]=~& \frac{1}{2}\sum\limits_{h',g'=0,1}{(-)^{\frac{1-\epsilon}{2}(hg'-gh')} Z[^{h,h'}_{g,g'}]}  \\ \nonumber
=&\left[\frac{1}{2}\sum\limits_{\bar{a},\bar{b}=0,1}{(-)^{\bar{a}+\bar{b}+\bar{a}\bar{b}}~C_1~\frac{\bar\theta[^{\bar{a}}_{\bar{b}}]^2\bar\theta[^{\bar{a}+h}_{\bar{b}+g}]\bar\theta[^{\bar{a}-h}_{\bar{b}-g}]}{\bar\eta^4} }\right]~\Gamma_{(4,4)}[^h_g]\\ \label{Z2primeWL_a}
\times&\tilde{\Gamma}_{(1,1)}[^{X}_{Y}]\left(R/2\right)~\left[\frac{1}{2}\sum\limits_{k,\ell=0,1}{C_2~\frac{\theta[^k_\ell]^{6}\theta[^{k+h}_{\ell+g}]\theta[^{k-h}_{\ell-g}]}{\eta^8}}\right]~\left[\frac{1}{2}\sum\limits_{\rho,\sigma=0,1}{\frac{\theta[^\rho_\sigma]^{8}}{\eta^8}}\right].
\end{align}
Here,
\begin{align}\label{Z2primeWL_b}
	\tilde{\Gamma}_{(1,1)}[^{X}_{Y}]\left(R/2\right) = \frac{R/2}{\sqrt{2\tau_2}}~\sum\limits_{\tilde{m},n\in\mathbb{Z}}{e^{-\frac{\pi}{2\tau_2}(R/2)^2\left|\tilde{m}+\tau n\right|^2}(-)^{\tilde{m}X+nY}},
\end{align}
and
\begin{align}\label{Z2primeWL_c}
		X ~\equiv~k+\rho+\left(\frac{1-\epsilon}{2}\right)h ,\\
		Y ~\equiv~\ell+\sigma+\left(\frac{1+\epsilon}{2}\right)g .
\end{align}
Note that the `shift' parameters $X,Y$ now depend on the $\epsilon$-parameter that was previously 
introduced as a discrete torsion and which will be shortly reinterpreted as a parameter controlling the choice of Wilson line along the $S^1$.

In order to illustrate the effect that the $X,Y$-coupling of the $(1,1)$-lattice to the gauge charges has on the spectrum, we Poisson resum the $\tilde{\Gamma}_{(1,1)}[^X_Y]$-lattice and cast it in Hamiltonian form:
\begin{align}\label{Z2primeWL_d}
	\tilde{\Gamma}_{(1,1)}[^X_Y](R/2) = \sum\limits_{m,n\in\mathbb{Z}}{ (-)^{nY}\Lambda_{2m+X,\frac{n}{2}}(R)}.
\end{align}
From this form, it becomes clear that the $\mathbb{Z}_2'$-freely acting orbifold has the effect of
 shifting the momentum quantum number by $\frac{1}{2}X$ and also modifying the generalized GSO- or $\mathbb{Z}_2$-orbifold projections depending on the winding of the states.

Let us now comment on how Spinor-Vector duality arises in this framework. We focus only on the
 twisted sector $h=1$ and notice that only states with $\rho=0$, i.e. states which are ``uncharged" under $SO(16)$, can contribute to the massless spectrum.

Let us pick directly the vectorial of $SO(12)$, by noticing that if it exists in the massless spectrum it must necessarily come from the sector:
\begin{align}
		P_o \Gamma^{h=1}_{(+)} \times \Lambda_{2m+\frac{1-\epsilon}{2},n}(R)\times V_{12}\times\{S_4\oplus C_4\}\times \{O_{16}\oplus V_{16}\}.
\end{align}
A few comments are in order here. First of all, after performing the $\sigma$-projection, one finds that only the vacuum 
representation $O_{16}$ of $SO(16)$ survives. Of course, massless states can only occur in the sector of unshifted 
momentum/winding quantum numbers, which permits us to restrict our attention only to states with $X\in 2\mathbb{Z}$. For the vectorial representation, $X=(1-\epsilon)/2$, so that the vectorial can become massless 
only for the choice $\epsilon=+1$, as found in the previous section. In addition, the $P_o$ representation 
carries conformal weight $(0,\frac{1}{4})$, while the low-lying modes in $P_v$ start from conformal weight $(0,\frac{3}{4})$ and 
are already anti-chirally massive. Therefore, since the vectorial representation $V_{12}$ has conformal weight $(\frac{1}{2},0)$, only states from the $P_o \Gamma_{(+)}$-sector can be massless, because the 
contribution of the twisted lattices to the conformal weights are $(\frac{1}{4},\frac{1}{4})$ for $\Gamma_{(+)}^{h=1}$ 
and $(\frac{3}{4},\frac{1}{4})$ for $\Gamma_{(-)}^{h=1}$. Furthermore, the generalized GSO-projection realized by 
the $\ell$-summation selects the $C_4$ representation of $SO(4)$, whereas the orbifold projection $g'$ projects onto 
states where the twisted $\{\Gamma_{(\pm)}^{h=1}\}$-lattices and the twisted $\{P_o, P_v\}$ characters are 
correlated with the same $\mathbb{Z}_2$-parity, so that the surviving representation is :
\begin{align}\label{V12}
		P_o \Gamma^{h=1}_{(+)} \times \Lambda_{2m+\frac{1-\epsilon}{2},n}(R)\times V_{12}C_4 O_{16}.
\end{align}

In the same spirit, we can construct the massless states in the vacuum representation of $SO(12)$:
\begin{align}\label{O12}
		P_o \Gamma^{h=1}_{(-)} \times \Lambda_{2m+\frac{1-\epsilon}{2},n}(R)\times O_{12} S_4 O_{16}.
\end{align}
This time the $\ell$-projection will pick the $S_4$ representation of $SO(4)$, while the balance of 
conformal weights now indicates that the twisted lattice $\Gamma_{(-)}^{h=1}$ with odd $\mathbb{Z}_2$-parity has 
to be used in order for these states to become massless. As this representation comes from the same 
sector $k=\rho=0$ as the vectorial, the conditions for it to be massless are again $\epsilon=+1$, so that 
this singlet representation of $SO(12)$ is always present in the massless spectrum whenever the vectorial one is.

Similarly, we can construct the spinorial of $SO(12)$ as:
\begin{align}\label{S12}
		P_o \Gamma^{h=1}_{(+)} \times \Lambda_{2m+\frac{1+\epsilon}{2},n}(R)\times S_{12} \{O_4+V_4\} O_{16}.
\end{align}
Again, the $\ell$-projection picks the vacuum representation of $SO(4)$ and the balance of 
conformal weights straightforwardly picks the $\Gamma_{(+)}^{h=1}$ twisted lattice. Then, the $\mathbb{Z}_2$-projection 
can be carried out to show that this representation is indeed invariant. The condition for it to be massless is again, $X\in 2\mathbb{Z}$, which is now satisfied for $\epsilon=-1$. 

This reproduces exactly the same conditions for the Spinor-Vector duality of the previous section, in 
terms of the choice of the discrete torsion. This simple and direct check will be generalized in the next 
section to provide the conditions for the presence of the Spinor-Vector duality, in terms of the value of the Wilson line along $S^1$.


\subsection{Turning on a general Wilson line background}\label{GeneralWL}

Since the effect of the freely-acting $\mathbb{Z}_2'$ is simply to correlate half-shifts along the $S^1$-circle with the charges of the 
gauge sector, it must have a natural interpretation as a special Wilson line. We will illustrate this point and identify the 
particular choices of Wilson lines that correspond to the two possible values of the $\epsilon$-parameter.
 To this end, we will start with the initial $\mathcal{N}_4=2$ theory compactified on $S^1\times\tilde{S}^1\times T^4/\mathbb{Z}_2$, 
before the freely-acting $\mathbb{Z}_2'$ orbifold is introduced. The partition function is again simply the sum of the $\mathbb{Z}_2$ orbifold blocks as in (\ref{Z2orbBlocks}), where now:
\begin{align}\nonumber
		Z[^h_g] =& \left[\frac{1}{2}\sum\limits_{\bar{a},\bar{b}=0,1}{(-)^{\bar{a}+\bar{b}+\bar{a}\bar{b}}~C_1~\frac{\bar\theta[^{\bar{a}}_{\bar{b}}]^2\bar\theta[^{\bar{a}+h}_{\bar{b}+g}]\bar\theta[^{\bar{a}-h}_{\bar{b}-g}]}{\bar\eta^4} }\right]~\Gamma_{(4,4)}[^h_g]\\ \label{N=2_SelfDual}
\times&\Gamma_{(1,1)}(R/2)~\left[\frac{1}{2}\sum\limits_{k,\ell=0,1}{C_2~\frac{\theta[^k_\ell]^{6}\theta[^{k+h}_{\ell+g}]\theta[^{k-h}_{\ell-g}]}{\eta^8}}\right]~\left[\frac{1}{2}\sum\limits_{\rho,\sigma=0,1}{\frac{\theta[^\rho_\sigma]^{8}}{\eta^8}}\right],
\end{align}
and here the $\Gamma_{(1,1)}$-lattice is a spectator:
$$
	\Gamma_{(1,1)}(R/2)= \frac{R/2}{\sqrt{2\tau_2}}~\sum\limits_{\tilde{m},n\in\mathbb{Z}}{e^{-\frac{\pi }{2\tau_2}(R/2)^2\left|\tilde{m}+\tau n\right|^2}}.
$$
We will now turn on a general Wilson line along the $X^8$ compact direction. This amounts to a perturbation
of the $\sigma$-model by the injection of the following marginal $(1,1)$-operator:
\begin{align}\label{WLdeform}
		\delta S = \int{d^2 z~A_8^a\,\bar\partial X^8(\bar{z}) J^a(z)},
\end{align}
$J^a(z)$, with $a=1,\ldots, 16$, are the currents in the Cartan subalgebra of $E_8\times E_8$. In the 
fermionic formulation they correspond to left-moving complex fermion bilinears $J^a(z)=i\bar\Psi^a\Psi^a(z)$. This deformation corresponds to
 turning on a non-trivial Wilson line $A^a_8$ around the $S^1$-circle. In fact, it is more convenient to factor out the $S^1$ radius and define $y^a \equiv A_8^a/R$. The effect of the insertion of this 
marginal operator is to deform the torus amplitude. By modifying the boundary conditions of the left-moving fermions, or by directly carrying out the path integral one finds:
\begin{align}\nonumber
			Z&[^h_g] = \\ \nonumber
&\left[\frac{1}{2}\sum\limits_{\bar{a},\bar{b}=0,1}{(-)^{\bar{a}+\bar{b}+\bar{a}\bar{b}}~C_1~\frac{\bar\theta[^{\bar{a}}_{\bar{b}}]^2\bar\theta[^{\bar{a}+h}_{\bar{b}+g}]\bar\theta[^{\bar{a}-h}_{\bar{b}-g}]}{\bar\eta^4} }\right]\Gamma_{(4,4)}[^h_g]~\left[~\sum\limits_{\tilde{m},n\in\mathbb{Z}}{ e^{-\frac{\pi}{\tau_2}(R/2)^2\left|\tilde{m}+\tau n\right|^2} }\right.
\\  &\times\left.\frac{1}{2}\sum\limits_{k,\ell=0,1}{ C_2\left(\frac{\theta[^{k+h-2y^1 n}_{\ell+g-2y^1\tilde{m}}]\theta[^{k-h-2y^2 n}_{\ell-g-2y^2\tilde{m}}] }{\eta^2}  \prod\limits_{B=3}^{8}{\frac{\theta[^{k-2y^B n}_{\ell-2y^B\tilde{m}}]}{\eta}} \right)} \frac{1}{2}\sum\limits_{\rho,\sigma=0,1}{\prod\limits_{C=9}^{16}{\frac{\theta[^{\rho-2y^C n}_{\sigma-2y^C\tilde{m}}]}{\eta}} }~e^{-i\pi\Xi_{\tilde{m},n}(y)}  \right] ,
\end{align}
where the phase:
\begin{align}
		\Xi_{\tilde{m},n}(y)= \tilde{m}n\sum\limits_{a=1}^{16}{y^a y^a} - n\left((\ell+g)y^1+(\ell-g)y^2+\ell\sum\limits_{B=3}^{8}{y^B}+\sigma\sum\limits_{C=9}^{16}{y^C}\right),
\end{align}
ensures that the deformation preserves modular invariance. 

So far, a particularly simple $\mathbb{Z}_2\times\mathbb{Z}_2'$-model was chosen in order to 
exhibit the basic structure of Spinor-Vector duality. Initially, it was defined in terms of 
the freely-acting $\mathbb{Z}_2'$-orbifold correlating the gauge charges with half-shifts along 
a compact $S^1$, performed on top of the non-freely acting $\mathbb{Z}_2$-rotation orbifold. 
In the previous section, it was shown that the effect of the freely-acting component could be 
resummed, in order to provide a realization of the same model solely in terms of the orbifold blocks of the non-freely acting $\mathbb{Z}_2$, eqs. (\ref{Z2primeWL_a})-(\ref{Z2primeWL_d}).

Now we are in the position to demonstrate that the effect of the freely acting $\mathbb{Z}_2'$ is 
equivalent to turning on a particular choice of Wilson line, along $S^1$. In fact, it will be instructive to determine the general conditions for the choice of the Wilson line which result in Spinor-Vector duality.

In general, non-rational values for the Wilson line will typically break the enhanced gauge group down to its Cartan factors. 
However, rational values of the Wilson line may preserve the enhancement. For example, turning on
 a Wilson line with rational values $y^a=p/q$, with $p<q$ being relatively prime integers, is equivalent 
to a freely-acting $\mathbb{Z}_{(1+p\,\textrm{mod}2)q}$ orbifold. For simplicity, and for the purposes of 
our discussion, it will be sufficient to restrict our attention to the $\mathbb{Z}_2$-case, namely to 
specific discrete points $y^a\in\mathbb{Z}$ along the -otherwise continuous- Wilson line.

Using the periodicity properties of $\theta$-functions, it is straightforward to show that the 
partition function reduces exactly to the form  (\ref{Z2primeWL_a}), (\ref{Z2primeWL_b}) of the $Z[^h_g]$-orbifold blocks, where now the $\tilde{\Gamma}_{(1,1)}[^X_Y]$-lattice of (\ref{Z2primeWL_c}) is shifted by:
\begin{align}
		X=& (k+h)y^1+(k-h)y^2+ k\sum\limits_{B=3}^{8}{y^B}+\rho\sum\limits_{C=9}^{16}{y^C}+n\sum\limits_{a=1}^{16}{y^a y^a} ,  \\
		Y=& (\ell+g)y^1+(\ell-g)y^2+\ell\sum\limits_{B=3}^{8}{y^B}+\sigma\sum\limits_{C=9}^{16}{y^C} .
\end{align}
Comparison of the above equation with (\ref{Z2primeWL_c}) explicitly illustrates the correspondence. As a particular example, consider turning on the Wilson line:
\begin{align}\label{WL+}
		y^a=(0,0|1,0,0,0,0,0|1,0,0,0,0,0,0,0).
\end{align}
This choice corresponds precisely to the $\mathbb{Z}_2\times\mathbb{Z}_2'$ orbifold with the particular choice $\epsilon=+1$ for the discrete torsion.
For the opposite discrete torsion $\epsilon=-1$, one may take, instead:
\begin{align}\label{WL-}
	y^a = (1,0|0,0,0,0,0,0|1,0,0,0,0,0,0,0).
\end{align}
This clearly illustrates that the freely-acting $\mathbb{Z}_2'$ introduced in the original formulation of the model is nothing but a particular choice of the Wilson line around the compact $X^8$-circle.

Furthermore, it is easy to obtain the general conditions for the Wilson line $y^a\in\mathbb{Z}$ that lead 
to manifestations of Spinor-Vector duality. In particular, it has been already argued that the vectorial, 
vacuum and spinorial representations of $SO(12)$ arise from the sectors exhibited in eqs. (\ref{V12}), (\ref{O12}) and (\ref{S12}), 
respectively. Of course, since massless states come from the even winding sector, the $Y$-shift introduces no modification to the projections. 

The conditions for the low-lying states in these sectors to be massless can be found by imposing $X \in 2\mathbb{Z}$. 
Noting that, for these particular sectors, $h=1$ and $\rho=0$ we can distinguish between two conditions controlling the presence of massless states:
\begin{itemize}
	\item If $\sum\limits_{A=1}^{2}{y^A} \in 2\mathbb{Z}$, then both $k=0$-sectors $V_{12}$ and $O_{12}$ are massless.
	\item If $\sum\limits_{B=3}^{8}{y^B} \in 2\mathbb{Z}$, then the spinorial sector $S_{12}$ is massless.
\end{itemize}
Clearly, for certain choices of the Wilson line both conditions can be simultaneously satisfied, in which case 
one recovers the Spinor-Vector self-dual models with enhanced $E_7$-gauge symmetry. Similarly, it is possible 
to choose the Wilson line such that none of the above conditions are satisfied, in which case all charged hypermultiplets
 become massive. Finally, by choosing to violate only one out of the two conditions, the Wilson line higgses either 
the vectorial $V_{12}$ (always followed by the vacuum $O_{12}$) representations while keeping the spinorial $S_{12}$ massless, or vice-versa.

Let us note here an important property, present in this class of models, where $\mathcal{N}_4=2$ supersymmetry is 
preserved by the orbifold action. As shown in eq. (\ref{WLdeform}), the Wilson line deformation arises from the 
injection into the $\sigma$-model of a marginal operator which survives the orbifold projection. It is clear that 
this $(1,1)$-operator is associated to a scalar in the physical massless spectrum of the theory. In particular, it 
corresponds to a modulus within the $\frac{SO(16+2,2)}{SO(16+2)\times SO(2)}$-factor of the full moduli space and can,
 thus, take continuous values. This provides the basis for the continuous connection of all the above vacua, in view of 
the fact that they can all be recovered for specific choices of the Wilson line $y^a$ around $S^1$. 

In fact, this can be used to illustrate the fact that Spinor-Vector duality is directly interrelated with the enhancement
 of the gauge symmetry at the Spinor-Vector self-dual point. Indeed, by the very structure of the massless representations
 and the above conditions, it is clear that at the self-dual points $X\in 2\mathbb{Z}$, independently of the values of $k$ or $h$. This guarantees that the mapping
\begin{align}
\left\{\begin{array}{c}
		P_o\Gamma_{(-)}^{h=1}O_{12}S_4 O_{16}\\
		P_o\Gamma_{(+)}^{h=1}V_{12}C_4 O_{16}\\
		P_o\Gamma_{(+)}^{h=1}S_{12}O_4 O_{16}\\
\end{array}\right\} \leftrightarrow
		\left\{\begin{array}{c}
		Q_v\Gamma_{(+)}^{h=0}O_{12}O_4 O_{16}\\
		Q_v\Gamma_{(+)}^{h=0}O_{12}O_4 S_{16}\\
		Q_v\Gamma_{(+)}^{h=0}C_{12}C_4 O_{16}\\
\end{array}\right\},
\end{align}
preserves conformal weights and is one to one. This illustrates how gauge symmetry enhancement translates into Spinor-Vector self-duality in the twisted massless spectrum. 

In this particular example, by continuously deforming away from the critical self-dual point along one of the flat 
directions $y^1$, $E_7\times SU(2)$ spontaneously breaks down to $SO(12)\times U(1)\times U(1)$. When one reaches 
$y^1=1$, the gauge symmetry becomes enhanced back to $SO(12)\times SO(4)$ with the states transforming with the vectorial 
representation of $SO(12)$ having acquired a mass, while keeping the spinorial massless. An alternative way to give mass 
to the vectorial of $SO(12)$ is by deforming only one of the two $SU(2)$-factors. This corresponds to deforming along the
 trajectory $y^1=y^2=\lambda$. As we move continuously away from $\lambda=0$, the $E_7\times SU(2)$ breaks 
down to $SO(12)\times U(1)\times SU(2)$ and $V_{12}$ becomes massive. At point $\lambda= 1/2$, the $U(1)$ gets enhanced so that one again recovers $SO(12)\times SO(4)$.

One may try to deform along a different flat direction in order to give mass to the spinorial while keeping the vectorial massless.
 However, since the massless $SO(12)$-spinorials in the twisted sector are always attached to the vaccum representations 
of $SO(4)\times E_8'$, i.e. $S_{12}O_4 O_{16}$, the only way to give them mass (while keeping the vectorial massless) 
is via non-vanishing expectation values for the Wilson line around $S^1$, associated to the $SO(12)$ factor. This 
inevitably breaks $SO(12)$ spontaneously to one of its subgroups. As an example, consider deforming along the $y^3$ flat 
direction, where $SO(12)$ spontaneously breaks down to $SO(10)\times U(1)$, until the point $y^3=1$ is reached. There, 
the $SO(10)\times U(1)$ gets enhanced back to $SO(12)$ but, now, the spinorial representation has become massive whereas 
the vectorial is kept massless. Of course, deforming along some generic flat direction may render both vectorial and spinorial representations massive.

The possibility of continuous interpolation between vacua with massless vectorials and massless spinorials of $SO(12)$ is 
special to the $\mathcal{N}_4=2$ case. In the case of $\mathbb{Z}_2\times\mathbb{Z}_2$ models, which preserve $\mathcal{N}_4=1$ 
supersymmetry, and where both $\mathbb{Z}_2$'s act as rotations on the full $T^6$-torus, the situation is substantially different. 
There, Wilson lines no longer correspond to marginal operators surviving the orbifold projections and, as a result, the associated 
deformation parameters are no longer continuous. Instead, the only allowed possibility would be to turn on discrete Wilson lines, 
as in \cite{cfkr}. In fact, a special property of $\mathbb{Z}_2\times\mathbb{Z}_2$ models with $\mathcal{N}_4=1$ supersymmetry is 
that their twisted sectors actually inherit the structure of the $\mathcal{N}_4=2$ theories. However, whereas in the $\mathcal{N}_4=2$ case 
that we are considering here the Wilson line deformation parameters $y^a$ can be continuously varied, in the $\mathcal{N}_4=1$ case they can 
only take discrete values and are, essentially, discrete remnants of the Wilson line deformations of the $\mathcal{N}_4=2$ theory.

Therefore, the interpretation of Spinor-Vector duality as the result of turning on specific Wilson lines that give masses either to 
vectorial or spinorial representations, with the initial (undeformed $y^a=0$) theory containing both representations in its massless 
spectrum, has two important consequences. First of all, it unifies a class of models, including the model considered in this paper as 
well as those presented in ref. \cite{aft}, and exhibits their common origin. Secondly and most importantly, it sheds light into the 
origin and nature of the duality and exhibits its close relation to the inherently stringy phenomenon of symmetry enhancement. This will 
be discussed in more detail in the next section, where it will be shown how Spinor-Vector duality is, in fact, a discrete remnant of the 
spectral flow of a spontaneously broken, left-moving, (global) extended superconformal algebra.


\section{Spinor-Vector duality from $N=2$ and $N=4$ Spectral-Flow}\label{SpFlow}

In the previous sections we illustrated how Spinor-Vector duality arises by turning on particular Wilson lines 
in a `parent' theory with enhanced gauge symmetry. In the particular example considered there, different choices 
for the Wilson line around the  $S^1$-circle resulted in massive vectorial or spinorial representations of the 
$SO(12)$-gauge group. Let us recall that all vacua containing massless vectorial representations also contained massless 
hypermultiplets, which were singlets under the $SO(12)$-group. What is more, the total number of massless states in the 
twisted sector, vectorials $V_{12}$ and singlets $O_{12}$ on the one hand, and spinorials $S_{12}$ on the other, was found to 
be the same. This equality is not a numerical coincidence and  the reason behind this matching lies in the presence of (at least)
 an unbroken global $N=4$ worldsheet superconformal symmetry, in the left-moving sector, associated to the gauge symmetry enhancement in the Spinor-Vector self-dual case. 

The presence of this unbroken $N=4$ algebra can be seen as an embedding\footnote{This embedding of the spin connection of 
Type II into the gauge connection of Heterotic theories is known as the Gepner map \cite{Gepner}.} of the $N=4$ worldsheet superconformal 
algebra of Type II theories into the bosonic (left-moving) sector of the Heterotic string. The presence of an unbroken $N=4$ SCFT 
introduces a spectral flow, which can be seen to transform the spinorial representations of $SO(12)$ into the vectorial (always followed
 by the scalar) representations and vice-versa. In the following subsections, we will display the way the mapping arises in 
the $\mathcal{N}_4=1$ and $\mathcal{N}_4=2$ cases\footnote{As will be discussed in detail in this section, in the $\mathcal{N}_4=1$ case, 
the enhancement arises from the presence of an $N=2$ SCFT, whose spectral flow induces the spinor-to-vector map.}, by explicitly constructing 
the spectral-flow operator and exhibiting its action on the vertex operators.


\subsection{Spinor -Vextor Duality and $N=2$ Spectral Flow in $\mathcal{N}_4=1$ Vacua}

We will first start with the simpler case of unbroken $\mathcal{N}_4=1$ spacetime supersymmetry and expand upon 
the analysis of ref.\cite{cfkr}. Spacetime supersymmetry requires the local right-moving $\hat{c}=6$, $N_R=1$ internal SCFT to
 become enhanced to $N_R=2$. Now consider the case where the left-moving internal CFT also becomes enhanced to a global $N_L=2$ SCFT (see, for example, \cite{BanksDixon}, \cite{EguchiTaormina} and references therein). 

This enhancement arises naturally via the Gepner map, as follows. Consider first the left-moving worldsheet 
degrees of freedom of a Type II theory with an enhanced $N_L=2$ global superconformal algebra. The vertex operators are generically proportional to:
\begin{align}
		e^{q\phi+is_0 H_0+is_1 H_1+ i\frac{Q}{\sqrt{3}} H},
\end{align}
where $q$ is the superghost charge (picture), $s_0$, $s_1$ are the $SO(1,3)$ helicity charges and $Q$ is the
 charge with respect to the $U(1)$ current $J(z)=i\sqrt{3}\,\partial H(z)$ of the internal $N_L=2$ SCFT. The currents generating spacetime supersymmetry are constructed in terms of the free boson as:
\begin{align}\nonumber
		& e^{-\phi/2}S_{\alpha}\Sigma(z)= e^{-\frac{1}{2}\phi \pm\frac{i}{2}(H_0+H_1) + i\frac{\sqrt{3}}{2}H}~,\\ \label{SUSYcurrent}
		& e^{-\phi/2}C_{\dot\alpha}\Sigma^{\dagger}(z)=  e^{-\frac{1}{2}\phi \pm\frac{i}{2}(H_0-H_1) - i\frac{\sqrt{3}}{2}H}~.
\end{align}
Here $\Sigma, \Sigma^\dagger$ are the maximal charge ground states of the R-sector with conformal weight $(\frac{3}{8}$,0). Imposing a good action of the supersymmetry currents on the vertex operators of the spectrum requires:
\begin{align}
	q+s_0+s_1+Q \in 2\mathbb{Z}~.
\end{align}
The $N_L=2$ spectral flow arises from shifting the $U(1)$ charges so that one may obtain a continuous interpolation between the NS ($\alpha=0$) and R ($\alpha=\pm\frac{1}{2}$) sectors:
\begin{align}\nonumber
	 & J_n \rightarrow J_n - 3\alpha \delta_{n,0}\\ \label{SFlow}
	 & L_n \rightarrow L_n - \alpha J_n + \frac{3}{2}\alpha^2 \delta_{n,0}.
\end{align} 

Now consider embedding the $N_L=2$ SCFT into the left-moving (`bosonic') sector of the $E_8\times E_8$ Heterotic string. The analogue of the supersymmetry current (\ref{SUSYcurrent}) is now built as:
\begin{align}\label{SFop}
		\mathcal{I}(z)\Sigma(z),
\end{align}
where $\mathcal{I}(z)$ is a $(\frac{5}{8},0)$-operator that dresses the internal $N_L=2$ ground state by replacing 
the superghost and spacetime fermion contributions $e^{-\phi/2\pm \frac{i}{2}H_0\pm\frac{i}{2}H_1}$ of the Type II case. In the 
Heterotic side this operator will arise from the gauge degrees of freedom. The spin connection may be naturally embedded into the gauge connection by setting:
\begin{align}
		\mathcal{I}(z)=e^{i \lambda\cdot Z(z)},
\end{align}
with $\lambda^A=\pm\frac{1}{2}$ and $A=1,\ldots 5$. This is simply the bosonization of the R-sector ground state for 
the 5 complex current algebra fermions $\Psi^A$ in $E_8$. The GSO projection is then naturally generalized by requiring that the spectral-flow operator (\ref{SFop}) has a well-defined action on the states:
\begin{align}\label{GSO}
		\sum\limits_{A=1}^{5}{\oint{\frac{dz}{2\pi i}\,\partial Z^A(z)}} + Q \in 2\mathbb{Z}.
\end{align}
This constrains the sum of the number operator for the $5$ current algebra fermions $\Psi^A$ and the charge $Q$ of the $N_L=2$ SCFT to be even.

The $(1,0)$-currents surviving the GSO projection are then:
\begin{align}
		\Psi^A \Psi^B(z)~~,~~S_{10}\Sigma(z)~~,~~C_{10}\Sigma^{\dagger}(z)~~,~~J(z).
\end{align}
Here $A,B=1,\ldots 5$ run over the $5$ complex current algebra fermions so that the fermion bilinears $\Psi^A\Psi^B$ transform 
as the adjoint representation $\textbf{45}$ of $SO(10)$. Similarly,  $S_{10}$ and $C_{10}$ are the R-sector vertex operators for 
the spinorial $\textbf{16}$ and conjugate spinorial $\overline{\textbf{16}}$ representations of $SO(10)$. Together with the singlet generated by $J(z)$, the currents form the adjoint representation $\textbf{78}$ of $E_6$.

As a concrete example, consider the $N=(2,2)$ compactification on $T^6/\mathbb{Z}_2\times\mathbb{Z}_2$, which 
preserves $\mathcal{N}_4=1$ left-moving spacetime supersymmetry. In terms of modular covariant conformal blocks, the above Gepner map is realized as:
\begin{align}
	\frac{1}{2}\sum\limits_{a,b}{(-)^{a+b+ab}\theta[^a_b]\theta[^{a+h_1}_{b+g_1}]\theta[^{a+h_2}_{b+g_2}]\theta[^{a-h_1-h_2}_{b-g_1-g_2}]}
 \rightarrow \frac{1}{2}\sum\limits_{k,\ell}{\theta[^k_\ell]^5\theta[^{k+h_1}_{\ell+g_1}]\theta[^{k+h_2}_{\ell+g_2}]\theta[^{k-h_1-h_2}_{\ell-g_1-g_2}]}.
\end{align}

The full modular invariant partition function of the theory can be organized into orbifold blocks, as before:
\begin{align}\nonumber
		Z[^{h_1,h_2}_{g_1,g_2}] =& \left[\frac{1}{2}\sum\limits_{\bar{a},\bar{b}=0,1}{(-)^{\bar{a}+\bar{b}+\bar{a}\bar{b}}
C_1~\frac{\bar\theta[^{\bar{a}}_{\bar{b}}]\bar\theta[^{\bar{a}+h_1}_{\bar{b}+g_1}]\bar\theta[^{\bar{a}+h_2}_{\bar{b}+g_2}]\bar\theta[^{\bar{a}-h_1-h_2}_{\bar{b}-g_1-g_2}]}{\bar\eta^4} }\right]
\Gamma_{(6,6)}[^{h_1,h_2}_{g_1,g_2}]\\ \label{N=1_Example}
\times&\left[\frac{1}{2}\sum\limits_{k,\ell=0,1}{C_2~\frac{\theta[^k_\ell]^{5}\theta[^{k+h_1}_{\ell+g_1}]\theta[^{k+h_2}_{\ell+g_2}]
\theta[^{k-h_1-h_2}_{\ell-g_1-g_2}]}{\eta^8}}\right]~\left[\frac{1}{2}\sum\limits_{\rho,\sigma=0,1}{\frac{\theta[^\rho_\sigma]^{8}}{\eta^8}}\right],
\end{align}
where, for concreteness, we make the following choice of chiralities:
\begin{align}\nonumber
	&C_1 = (-)^{\bar{a}\bar{b}},\\
	&C_2 = 1.
\end{align}
We will consider here the simple case where the twisted $\Gamma_{(6,6)}[^{h_1,h_2}_{g_1,g_2}]$ lattice is factorized into the product of three $\Gamma_{(2,2)}[^{h}_{g}]$ lattices as follows:
\begin{align}\label{LatticeFactorizes}
		\Gamma_{(6,6)}[^{h_1,h_2}_{g_1,g_2}] = \Gamma_{(2,2)}[^{h_1}_{g_2}]\Gamma_{(2,2)}[^{h_2}_{g_2}]\Gamma_{(2,2)}[^{h_1+h_2}_{g_1+g_2}],
\end{align}
where :
\begin{align}\label{LatticeFactorizes2}
		\Gamma_{(2,2)}[^{h}_{g}]=\left\{\begin{array}{c l}
	\Gamma_{(2,2)} &,~\textrm{for}~(h,g)=(0,0) \\
	\left|\frac{2\eta}{\theta[^{1-h}_{1-g}]}\right|^{2} &,~\textrm{for}~(h,g)\neq(0,0)\\
\end{array}\right.
\end{align}

In terms of the fermionic formulation of the Heterotic string, the $\hat{c}=6$ system realizing the $N_L=2$ SCFT, is built
 out of the $3$ complex fermions $\Psi^{6,7,8}$ which are bosonized as $e^{\pm iH^j(z)}$, with $j=6,7,8$. The spectral-flow currents are then constructed out of free fields as:
\begin{align}\nonumber
		C_{10}(z) e^{\frac{i}{2}H^6(z)+\frac{i}{2} H^7(z)+\frac{i}{2} H^8(z)} ~&\propto~ e^{i\frac{\sqrt{3}}{2} H(z)} ,\\ \label{SFop2}
		S_{10}(z) e^{-\frac{i}{2}H^6(z)-\frac{i}{2} H^7(z)-\frac{i}{2} H^8(z)} ~&\propto~ e^{-i\frac{\sqrt{3}}{2} H(z)} .
\end{align}
The presence of an $N_L=2$ SCFT manifests itself in the ability to factor out the free $U(1)$ current $J(z)$. To see this, consider the following linear field redefinition for the free scalars:
\begin{align}\nonumber
	H(z) &= \left( H^6+H^7+H^8 \right)/\sqrt{3} \\ \nonumber
	X(z) &= \left( 2H^6-H^7-H^8 \right)/\sqrt{6} \\
	Y(z) &= \left( - H^7 + H^8 \right)/\sqrt{2}.
\end{align}
The bosons $H,X,Y$ are still free and, in particular, $J(z)=i\sqrt{3}\,\partial H(z)$ is identified with the conserved $U(1)$ current of the $N_L=2$ SCFT, in accordance with (\ref{SFop}) and (\ref{SFop2}). 

Consider now the action of the spectral flow in the untwisted sector. The scalar spectrum contains states in the NS-sector 
of the $N=(2,2)$ SCFT saturating the BPS bound $(\Delta,\bar\Delta)=(\frac{1}{2}|Q|,\frac{1}{2}|\bar{Q}|)$. Their vertex operators 
can be written in terms of the $N=(2,2)$ chiral primaries $\mathcal{F}$ with charge $Q=\bar{Q}=1$ as:
\begin{align}\label{v10v2}
	\Psi^A(z) \mathcal{F}(z,\bar{z}) = \Psi^A(z) e^{ iH^j(z)} e^{-\bar\phi(\bar{z})+ i\bar{H}_k(\bar{z}) } ~\propto~e^{i\frac{1}{\sqrt{3}}H(z)},
\end{align}
where $A=1,\ldots,5$, $j=6,7,8$ and $k=2,3,4$. This transforms as the vectorial of $SO(10)$. Under the 
spectral flow (\ref{SFlow}), the left-moving $U(1)$ charge is shifted by $-\frac{3}{2}$ units to yield $Q=1 \rightarrow -\frac{1}{2}$. Taking, for example, $j=6$:
\begin{align}
		e^{\pm iH^6}~\rightarrow~ e^{\pm\frac{i}{2}H^6\mp\frac{i}{2}H^7\mp\frac{i}{2}H^8},
\end{align}
we see that the spectral flow transforms the $V_{10}V_{2}O_2 O_2$ representation (\ref{v10v2}) into 
the $C_{10}C_2 S_2 S_2$ and $S_{10}S_{2}C_2 C_2$ representations in the R-sector of the $N_L=2$ SCFT. It is described by the vertex operator:
\begin{align}\nonumber
	C_{10}(z) e^{\frac{i}{2}H^6(z)-\frac{i}{2}H^7(z)-\frac{i}{2}H^8(z)} e^{-\bar\phi(\bar{z})+ i\bar{H}_2(\bar{z}) } ~\propto~e^{-i\frac{1}{2\sqrt{3}}H(z)},\\ \label{c10c2}
	S_{10}(z) e^{-\frac{i}{2}H^6(z)+\frac{i}{2}H^7(z)+\frac{i}{2}H^8(z)} e^{-\bar\phi(\bar{z})+ i\bar{H}_2(\bar{z}) } ~\propto~e^{+i\frac{1}{2\sqrt{3}}H(z)},
\end{align}
transforming in the spinorial representation of $SO(10)$ and its conjugate, respectively. Note that the spectral 
flow between the NS and R sectors arises explicitly through the action of the $Q=\pm\frac{3}{2}$ operators in (\ref{SFop2}) on the states, in complete analogy to the case of spacetime supersymmetry.

Shifting the $U(1)$ charge of (\ref{c10c2}) once more, $Q=-\frac{1}{2}\rightarrow -2$, one finds the 
singlet representation $O_{10}O_2 V_2 V_2$ in the NS-sector of the $N=(2,2)$ SCFT. Its vertex operator is written in terms of the chiral primary $\mathcal{G}$ with charge $(Q,\bar{Q})=(-2,1)$:
\begin{align}\label{o10o2}
	\mathcal{G}(z,\bar{z}) = e^{-iH^7(z)-iH^8(z)}e^{-\bar\phi(\bar{z})+ i\bar{H}_2(\bar{z})} ~\propto~e^{-i\frac{2}{\sqrt{3}}H(z)}.
\end{align}

The same analysis can be carried out in the twisted sectors. For concreteness, consider the $(h_1,h_2)=(1,0)$ sector. 
The massless matter spectrum contains again fermionic states transforming in the vectorial representation of $SO(10)$. They are built out of twisted $Q=1$ chiral primaries:
\begin{align}\label{v10c2o2c2}
		\Psi^A(z) e^{\frac{i}{2}H^6(z)+\frac{i}{2}H^8(z)}
\Gamma_{(+,+,+)}^{(1,0)}(z,\bar{z}) e^{-\frac{1}{2}\bar\phi(\bar{z})+\frac{i}{2}\bar{H}_0(\bar{z}) +\frac{i}{2}\bar{H}_1(\bar{z})+\frac{i}{2}\bar{H}_3(\bar{z}) } ~\propto~e^{i\frac{1}{\sqrt{3}}H(z)}.
\end{align}
Here $\Gamma_{(+,+,+)}^{(1,0)}(z,\bar{z})$ is the weight-$(\frac{1}{4},\frac{1}{4})$ invariant twist-field, associated to the $(h_1,h_2)=(1,0)$-twisted $\Gamma_{(6,6)}$-lattice
\begin{align}\label{twistF}
		\Gamma^{(h_1,h_2)}_{(s_1 s_3,s_2 s_3)}= \Gamma^{h_1}_{(s_1)}\Gamma^{h_2}_{(s_2)}\Gamma^{h_1+h_2}_{(s_3)},
\end{align}
with $s_i=\pm$ being the definite $\mathbb{Z}_2$-parities of the three $(2,2)$-sublattices, 
defined by analogy to (\ref{twistLattice_a}). Note that the untwisted $\Gamma_{(-)}^{h=0}$ lattice with negative
 parity projects out the low-lying states, whereas $\Gamma^{h=0}_{(+)}$ preserves these states but starts with $(0,0)$-conformal dimension. On the other hand, each $\Gamma_{(2,2)}$-twisted lattice $\Gamma^{h=1}_{(+)}$ of 
positive parity has conformal weight $(\frac{1}{8},\frac{1}{8})$, while $\Gamma^{h=1}_{(-)}$ contains sectors with weights $(\frac{5}{8},\frac{1}{8})$ and $(\frac{1}{8},\frac{5}{8})$.

In terms of characters, (\ref{v10c2o2c2}) corresponds to the $V_{10}C_2 O_2 C_2$ representation. 
Under the spectral flow\footnote{Of course, by shifting the $U(1)$ charge by $+\frac{3}{2}$ units, 
the vectorial representation would be mapped into the massive spinorial, $S_{10}V_2 C_2 V_2$.}, it will
 be mapped into the conjugate spinorial representation $C_{10}O_2 S_2 O_2$ with charge $Q=-\frac{1}{2}$:
\begin{align}\label{c10o2s2o2}
		C_{10}(z) e^{-\frac{i}{2}H^7(z)}\Gamma_{(+,+,+)}^{(1,0)}(z,\bar{z}) e^{-\frac{1}{2}\bar\phi(\bar{z})+\frac{i}{2}\bar{H}_0(\bar{z})+\frac{i}{2}\bar{H}_1(\bar{z})+\frac{i}{2}\bar{H}_3(\bar{z}) } ~\propto~e^{-i\frac{1}{2\sqrt{3}}H(z)}.
\end{align}
To see this, we will explicitly construct the twisted spectral-flow operator with charge $Q=-\frac{3}{2}$ and
 consider its action on the vertex operator (\ref{v10c2o2c2}). Here, because the twist is only $\mathbb{Z}_2$, it is 
possible to represent the twist-field vertex operators $\Gamma_{(\pm)}^{h_i}(z,\bar{z})$ associated to the twisted 
lattice in terms of level-one free-fermion characters. To this end, note that the topological contribution $\Gamma_{(\pm)}^{h}(z,\bar{z})$ of the 
twisted lattice (\ref{LatticeFactorizes}),(\ref{LatticeFactorizes2}) can be represented in terms of free-fermions as:
\begin{align}
		\Gamma_{(s)}^{h=1} = \frac{1}{2^2 \eta^2\bar\eta^2}
\sum\limits_{g=0,1}\sum\limits_{\gamma,\delta=0,1}{(-)^{\left(\frac{1-s}{2}\right)g} \theta[^\gamma_\delta]\theta[^{\gamma+1}_{\delta+g}]\times\bar\theta[^\gamma_\delta]\bar\theta[^{\gamma+1}_{\delta+g}]}.
\end{align}
Performing the $\gamma$-summations and imposing the $\delta,g$-projections we find the explicit form of the free-fermion representations of the twisted vertex operators:
\begin{align}\nonumber
		\Gamma^{h=1}_{(+)}(z,\bar{z})~=~& \left\{~ O_2 S_2 \bar{O}_2\bar{S}_2 ~\oplus~ O_2 C_2 \bar{O}_2\bar{C}_2 ~\oplus~ V_2 S_2 \bar{V}_2\bar{S}_2 ~\oplus~ V_2 C_2 \bar{V}_2\bar{C}_2 \phantom{\frac{1}{2}}\right.\\ \label{twistFplus}
		& \left. \phantom{\frac{1}{2}} \oplus~ S_2 O_2 \bar{S}_2\bar{O}_2 ~\oplus~ S_2 V_2 \bar{S}_2\bar{V}_2 ~\oplus~ C_2 O_2 \bar{C}_2\bar{O}_2 ~\oplus~ C_2 V_2 \bar{C}_2\bar{V}_2 ~\right\}.
\end{align}
\begin{align}\nonumber
		\Gamma^{h=1}_{(-)}(z,\bar{z})~=~& \left\{~O_2 S_2 \bar{V}_2\bar{C}_2 ~\oplus~ O_2 C_2 \bar{V}_2\bar{S}_2 ~\oplus~ V_2 S_2 \bar{O}_2\bar{C}_2 ~\oplus~ V_2 C_2 \bar{O}_2\bar{S}_2 \phantom{\frac{1}{2}}\right. \\ \label{twistFminus}
		&\left. \phantom{\frac{1}{2}} \oplus~ S_2 O_2 \bar{C}_2\bar{V}_2 ~\oplus~ S_2 V_2 \bar{C}_2\bar{O}_2 ~\oplus~ C_2 O_2 \bar{S}_2\bar{V}_2 ~\oplus~ C_2 V_2 \bar{S}_2\bar{O}_2 ~\right\}.
\end{align}
For simplicity, we suppress the indices and directly label each vertex
 operator by its $SO(2n)$-representation so that, for example, the $O_2$-representation contains 
the identity operator $\textbf{1}_2(z)$ as its ground state and the adjoint representation (realized as a bifermion) at the first excited level.

In the twisted sectors, the spectral-flow currents (\ref{SFop2}) become extended by a \\
\emph{chiral} operator $\Omega_{(\pm,\pm,\pm)}(z)$,  acting on the twist-field contribution $\Gamma^{(1,0)}(z,\bar{z})$ associated 
to the twisted lattices. The total invariant spectral-flow current is then decomposed into the following contributions:
\begin{align}\nonumber
	j_{\textrm{s.f.}}(z)~=~	(C_{10}C_2 C_2 C_2)(z) \Omega_{(+,+,+)}(z) 
	~&\oplus~ (C_{10}C_2 S_2 S_2)(z) \Omega_{(-,+,+)}(z)\\ \nonumber
	~\oplus~ (C_{10}S_2 C_2 S_2)(z) \Omega_{(-,+,-)}(z)
	~&\oplus~ (C_{10}S_2 S_2 C_2)(z) \Omega_{(-,+,-)}(z)\\ \nonumber
	~\oplus~ (S_{10}S_2 S_2 S_2)(z) \Omega_{(+,+,+)}(z)
	~&\oplus~ (S_{10}S_2 C_2 C_2)(z) \Omega_{(-,+,+)}(z)\\ \label{SFcurrent}
	~\oplus~ (S_{10}C_2 S_2 C_2)(z) \Omega_{(-,+,-)}(z)
	~&\oplus~ (S_{10}C_2 C_2 S_2)(z) \Omega_{(+,+,-)}(z) .
\end{align}
The chiral dressing $\Omega_{(\alpha,\beta,\gamma)}(z)$ transforms as $(\alpha\gamma , \beta\gamma)$ under the action of the 
orbifold group $\mathbb{Z}_2\times\mathbb{Z}_2$, where $\alpha,\beta,\gamma = \pm 1$. Its conformal weight 
is $( \frac{3-\alpha-\beta-\gamma}{2},0)$ and its action on the relevant twisted vertex operators of the form (\ref{twistF}), (\ref{twistFplus}), (\ref{twistFminus}) follows the fusion rule:
\begin{align}
		\Omega_{(\alpha,+,\gamma)}(z) ~\cdot~\Gamma^{(1,0)}_{(r,+,s)}(w,\bar{w}) = \frac{\Gamma^{1}_{(\alpha r)}\Gamma^{0}_{(+)}\Gamma^{1}_{(\gamma s)}(w,\bar{w})}{(z-w)^{\frac{1}{2}-\frac{1}{4}(\alpha+\gamma)}}+\ldots,
\end{align}
where again, we use a compact notation where the representation indices as well as the
 associated Dirac matrices arising from the OPEs are suppressed. The ellipsis denotes less singular terms. In terms of free bosonic fields, $\Omega_{(\alpha,\beta,\gamma)}(z)$ can be represented as:
\begin{align}
		\Omega_{(\alpha,\beta,\gamma)}(z) = e^{i\frac{1-\alpha}{2}(\pm \Phi_1\pm \Phi_2)+i\frac{1-\beta}{2}(\pm \Phi_3\pm \Phi_4)+i\frac{1-\gamma}{2}(\pm \Phi_5\pm \Phi_6)}.
\end{align}

We are now ready to explicitly calculate the action of the spectral-flow operator on the vertex 
operators. Let us start with $C_{10} O_2 S_2 O_2$ given in (\ref{c10o2s2o2}). The spectral-flow operator responsible for the mapping between the various representations is the zero mode of the current (\ref{SFcurrent}) :
\begin{align}\nonumber
		Q_{\textrm{s.f.}}=&\oint\limits{}\frac{dz}{2\pi i}\left[\phantom{\frac{}{}}\right.(C_{10}C_2 C_2 C_2)(z) \Omega_{(+,+,+)}(z) 
	~+~ z(C_{10}C_2 S_2 S_2)(z) \Omega_{(-,+,+)}(z)  \\ \nonumber
	~&+~ z^2(C_{10}S_2 C_2 S_2)(z) \Omega_{(-,+,-)}(z)
	~+~ z^2(C_{10}S_2 S_2 C_2)(z) \Omega_{(+,+,-)}(z)\\ \nonumber
	~&+~ (S_{10}S_2 S_2 S_2)(z) \Omega_{(+,+,+)}(z)
	~+~ z(S_{10}S_2 C_2 C_2)(z) \Omega_{(-,+,+)}(z)\\ \label{SFcharge}
	~&+~ z^2(S_{10}C_2 S_2 C_2)(z) \Omega_{(-,+,-)}(z)
	~+~ \left.z(S_{10}C_2 C_2 S_2)(z) \Omega_{(+,+,-)}(z) \phantom{\frac{}{}}\right].
\end{align}
Note that this operator could also be obtained as the invariant truncation of the 
untwisted spectral-flow current $C_{16}(z)$ of $E_8$. The lattice dressing $\Omega(z)$ ensures the operator 
survives the orbifold projections. The spectral flow charge can now act upon the various states $\mathcal{V}(0,0)|0\rangle$ 
centered at $z=\bar{z}=0$. In particular, its action on $C_{10}O_2 S_2 O_2$ generates the following (massless) representations:
\begin{align}\nonumber
		Q_{\textrm{s.f.}} \cdot &\left[C_{10}O_2 S_2 O_2~\Gamma^{(1,0)}_{(+,+,+)}e^{-\frac{1}{2}\bar\phi+\frac{i}{2}\bar{H}_0}\bar{S}_2\bar{O}_2\bar{S}_2\bar{O}_2\right](0,0)|0\rangle \\ \nonumber
		=~ &\left[~V_{10}C_2 O_2 C_2 \Gamma^{(1,0)}_{(+,+,+)}~+~ O_{10}S_2 O_2 C_2 \Gamma^{(1,0)}_{(-,+,+)} ~+~ O_{10}S_2 V_2 S_2 \Gamma^{(1,0)}_{(+,+,+)} \right. \\
		&~~+~ \left.O_{10}C_2 O_2 S_2 \Gamma^{(1,0)}_{(+,+,-)}\right] e^{-\frac{1}{2}\bar\phi+\frac{i}{2}\bar{H}_0}\bar{S}_2\bar{O}_2\bar{S}_2\bar{O}_2 |0\rangle .
\end{align}
It is straightforward to verify the above mapping by using the OPEs between $SO(N)$-spin fields, given in Appendix \ref{OPEs}.

As expected, the conjugate spinorial $SO(10)$-representation has been mapped into the vectorial $V_{10}C_2 O_2 C_2$, accompanied 
by the singlets. This illustrates the exact map between these representations, as it is induced by the spectral-flow current (\ref{SFcurrent}). It is instructive to count the numbers of (massless) degrees of freedom :
\begin{align}
		\begin{array}{l| l l}
				\textrm{Spinorial}~ & ~C_{10}O_2 S_2 O_2 & \rightarrow~~ 2^{5-1}\times (4\times 4) \\ \hline \hline
				\textrm{Vectorial}~ & ~V_{10}C_2 O_2 C_2 & \rightarrow~~ 10\times(4\times 4)\\ \hline
				\textrm{Singlets}~ & ~O_{10}S_2 O_2 C_2 & \rightarrow~~ 1\times(8\times 4)\\
				~ & ~O_{10}S_2 V_2 S_2 & \rightarrow~~ 2\times(4\times 4)\\
				~ & ~O_{10}C_2 O_2 S_2 & \rightarrow~~ 1\times(4\times 8)\\
		\end{array}
\end{align}

The spectral-flow displayed above gives rise to a number of supersymmetric-like identities, realized internally in 
the left-moving sector. Here, in contrast to the right-moving sector, which enjoys local worldsheet supersymmetry,
 the spinors and vectors transform under the gauge rotation group $SO(10)$ rather than the spacetime little group 
and, hence, there is no cancellation between the contributions of the vectorial and spinorial representations\footnote{Of course, 
this implementation of the correct spin-statistics arises automatically, from the requirements of higher-genus modular 
invariance and factorization.}. The analogues of ``bosons'' and ``fermions'' have again an equal contribution to the
 partition function, however, the characters are summed rather than subtracted. Nevertheless, these identities may 
lead to considerable simplifications in various calculations involving, for example, integration of the modular parameters
 over the fundamental domain. Such integrations arise frequently when one is calculating the one-loop vacuum amplitude in
 cases where the spacetime supersymmetry is spontaneously broken. 

For example, one may pick the weight $(*,\frac{1}{4})$-contribution of the internal CFT, which is relevant for 
creating massless states in the right-moving side. For concreteness, let us take the contribution that couples 
to the right-moving lattice piece $\bar{O}_2\bar{S}_2\bar{O}_2\bar{O}_2\bar{O}_2\bar{S}_2\in \Gamma^{(1,0)}_{(\pm,+,\pm)}$. Gathering 
together all relevant factors one may write this contribution as $\mathcal{U}(\epsilon)\bar{O}_2\bar{S}_2\bar{O}_2\bar{O}_2\bar{O}_2\bar{S}_2$, where:
\begin{align}
	\mathcal{U}(\epsilon)\equiv O_2 O_2\times \left[(O_2 S_2 O_2 S_2)A_1+(V_2 C_2 O_2 S_2)A_2+(O_2 S_2 V_2 C_2)A_3+(V_2 C_2 V_2 C_2)A_4\right],
\end{align}
and the $A_i$ are:
\begin{align}\nonumber
		&A_1 = \left[O_{10}S_2 V_2 S_2 +V_{10}C_2 O_2 C_2+\epsilon(C_{10}O_2 S_2 O_2+S_{10}V_2 C_2 V_2)\right] O_{16}\\ \nonumber
		&A_2 = \left[O_{10}S_2 O_2 C_2 +V_{10}C_2 V_2 S_2+\epsilon(C_{10}O_2 C_2 V_2+S_{10}V_2 S_2 O_2)\right] O_{16}\\ \nonumber
		&A_3 = \left[O_{10}C_2 O_2 S_2 +V_{10}S_2 V_2 C_2+\epsilon(C_{10}V_2 C_2 O_2+S_{10}O_2 S_2 V_2)\right] O_{16}\\
		&A_4 = \left[O_{10}C_2 V_2 C_2 +V_{10}S_2 O_2 S_2+\epsilon(C_{10}V_2 S_2 V_2+S_{10}O_2 C_2 O_2)\right] O_{16}
\end{align}
Here, $\epsilon=\pm 1$ denotes the spin-statistics sign. The value $\epsilon=-1$ would arise in the presence of
 worldsheet super-reparametrization invariance, hence, requiring a \emph{local} $N=1$ worldsheet SCFT. Its further 
enhancement to a global $N=2$ would introduce the spectral-flow responsible for $\mathcal{N}_4=1$ spacetime supersymmetry.
 The spacetime supersymmetric structure manifests itself in terms of chiral identities between current algebra characters,
 such as $\mathcal{U}(-1)=0$, which can be verified by using the Jacobi theta-function identities. 

In our case, worldsheet supersymmetry is global in the left-moving sector and there are no pictures, which corresponds 
to the `bosonic' case $\epsilon=+1$. Even though $\mathcal{U}(+1)$ is non-vanishing, the previous identity $\mathcal{U}(-1)=0$ may 
still be used to illustrate the spectral-flow and to algebraically simplify the characters. Examples of such identities will be 
presented in more detail in the next section, where the $\mathcal{N}_4=2$ case will be considered. However, they are still 
present in the $\mathcal{N}_4=1$ case as well, even though they are somewhat more tedius to display explicitly.

A very important observation can be made already at this point. It turns out that the spectral-flow operator in the twisted 
sector is none other than a deformed version of the operator inducing the \emph{Massive Spectral boson-fermion Degeneracy Symmetry} 
(MSDS) of \cite{MSDS}, \cite{ReducedMSDS}. The action of the MSDS-operator on states is only well-defined provided a set of conditions 
is satisfied, \cite{ReducedMSDS}, and these severely constrain the compactification. Whenever these are met, the chiral character identities
 emanating from the MSDS spectral-flow can be utilised to relate `vectorial' representations to `spinorials', with the exception of 
weight $\Delta=\frac{1}{2}$ ground states\footnote{States of conformal weight $(\frac{1}{2},*)$ are (chirally) massless in Type II
 theories. These are precisely the states that remain invariant under the MSDS spectral-flow, in contrast to the case of conventional 
supersymmetry. The principle governing the MSDS spectral-flow is preserved intact when the spin connection of Type II theories is embedded
 in the gauge connection of the Heterotic string.} which remain untransformed. In these cases, the spectral-flow operator precisely coincides 
with the MSDS charge, which is the zero mode of an invariant truncation of the $SO(24)$ spin-field $C_{24}(z)$. 

In the particular example (\ref{N=1_Example}), (\ref{LatticeFactorizes}) we considered in this section, 
the $\Gamma_{(6,6)}$-lattice in the $(h_1,h_2)=(1,0)$ plane was factorizable into three $\Gamma_{(2,2)}$ 
sublattices $\Gamma[^{~1~}_{\,g_1}]\Gamma[^{~0~}_{\,g_2}]\Gamma[^{~~1~~}_{g_1+g_2}]$. This particular compactification 
does not satisfy\footnote{It is possible to consider a discrete shift in the toroidal background parameters, compatible 
with the orbifold, such that the MSDS spectral-flow conditions would be satisfied in the $(1,0)$-plane.} the conditions 
for the MSDS spectral-flow and, as a result, identities such as those mentioned above can generically only arise in certain
 sub-sectors in which the spectral-flow current has a well-defined action. Of course, for the purposes of phenomenology, only
 the subsectors contributing to the massless spectrum are relevant. This is the case for the identity $\mathcal{U}(-1)=0$ considered 
above. Similar identities can be obtained by considering other contributions in $\Gamma^{(1,0)}_{(\pm,+,\pm)}$, such as those 
coupling to $\bar{O}_2\bar{S}_2\bar{O}_2\bar{O}_2\bar{O}_2\bar{C}_2$, or $\bar{O}_2\bar{C}_2\bar{O}_2\bar{O}_2\bar{O}_2\bar{C}_2$ and so on, 
provided that the right-moving lattice contribution has conformal weight $(0,\frac{1}{4})$ in order to produce anti-chirally massless states.

Finally, by turning on discrete torsions\footnote{These can be seen, alternatively, as discrete Wilson lines.},
 as in \cite{cfkr}, one may deform the theory and give masses to the vectorial representation (always accompanied by the singlets) or 
to the conjugate spinorial. Of course, this (discrete) deformation away from the extended symmetry point will have the effect of breaking
 the enhanced $N=(2,2)$ SCFT down to $N=(0,2)$, as is required for the preservation of the $\mathcal{N}_4=1$ spacetime supersymmetry. 

It is then clear that the observed duality map between the two theories, one with massless (conjugate) spinorials and one
 with massless vectorials, is the direct result of the spectral-flow in the twisted sectors of the enhanced $N=(2,2)$ compactification. 
In particular, the spectral-flow at the enhanced point guarantees that the number of massless degrees of freedom in the two -seemingly disconnected- theories always remains the same.


\subsection{Spinor-Vector Duality and $N=4$ Spectral Flow in $\mathcal{N}_4=2$ Vacua}

In this section, we will briefly extend the analysis of the previous section to the $\mathcal{N}_4=2$ level. 
This time, spacetime supersymmetry requires the extension of the local right-moving $N_R=1$, $\hat{c}=6$ superconformal 
system into a free $N_R=2$, $\hat{c}=2$ SCFT system and an $N_R=4$ SCFT with $\hat{c}=4$ \cite{BanksDixon}:
\begin{align}
		\{\,N=1\,,\,\hat{c}=6\,\} ~\longrightarrow~ \{\,N=2\,,\,\hat{c}=2\,\}\,\oplus\,\{\,N=4\,,\,\hat{c}=4\,\}.
\end{align}
As before, we are interested in the case where the left-moving internal CFT also becomes 
enhanced to a direct sum of global $\{N_L=4,\hat{c}=4\}\oplus\{N_L=2,\hat{c}=2\}$ SCFTs. In particular, the free $\hat{c}=2$ system will give rise to a compactification on $T^2$.

The starting point is, again, the Type II theory with $N_4=2$ supersymmetries arising from the left-moving side. The vertex operators of the states are now proportional to:
\begin{align}\label{states}
		e^{q\phi+i s_0 H_0+is_1 H_1+i r Y+iQ\sqrt{2}H},
\end{align}
where the spacetime part is defined as in the previous section, $r$ is the $U(1)$ charge of the 
free $j(z)=i\partial Y(z)$ boson and $Q$ is the `isospin' charge with respect to the diagonal $SU(2)_{k=1}$ 
current $J^3(z)=\frac{i}{\sqrt{2}}\partial H(z)$ of the internal $N_L=4$ SCFT. The two spacetime supersymmetry 
currents are then of the form (\ref{SUSYcurrent}), the difference now is the presence of two weight-$(\frac{3}{8},0)$ R-ground states $\Sigma^1(z)$, $\Sigma^2(z)$. In terms of the $Y(z)$,$H(z)$-scalars, they can be written as:
\begin{align}\nonumber
		&\Sigma^1(z)=e^{\frac{i}{2}Y(z)+i\frac{1}{\sqrt{2}}H(z)}, \\
		&\Sigma^2(z)=e^{\frac{i}{2}Y(z)-i\frac{1}{\sqrt{2}}H(z)}.
\end{align}
The fermionization of $Y(z)$ provides the 2 real fermions of the free $\hat{c}=2$ system. The generalization of the GSO projection ensuring the well-defined action of both supersymmetry currents on the states (\ref{states}) requires:
\begin{align}
		q+s_0+s_1+r+2Q \in 2\mathbb{Z}~~~,~~~	2Q \in \mathbb{Z}.
\end{align}
The $N_L=4$ spectral-flow is similarly \cite{EguchiTaormina}:
\begin{align}\nonumber
		&J_n^3\rightarrow J_n^3-\alpha\delta_{n,0}~,\\
		&L_n\rightarrow L_n-2\alpha J_n^3+\alpha^2\delta_{n,0}~.
\end{align}

We consider now the embedding of the left-moving spin connection of Type II into the `bosonic' sector of the Heterotic string.
The analysis is straightforward and parallel to the $N_L=2$ case of the previous section. To this end, we build the spectral flow current as in (\ref{SFop}).
We will extend the $SO(10)$ current algebra of complex fermions $\Psi^A$ (where $A=1,\ldots,5$) with the 
free complex fermion $\Psi^6(z)\equiv e^{iY(z)}$ of the $\hat{c}=2$ system so that $\Psi^A$ will be, henceforth,
 taken to generate an $SO(12)_{k=1}$ current algebra. Hence, the GSO projection of (\ref{GSO}) is carried through to the present case without modification. The invariant $(1,0)$-currents are then:
\begin{align}
		\Psi^A\Psi^B(z)~~,~~ C_{12} e^{\pm i\frac{1}{\sqrt{2}}H}(z)~~,~~J^3(z)~~,~~ J^{\pm}(z),
\end{align}
where $J^\pm(z)=e^{\pm i\sqrt{2}H(z)}$ and $I=1,\ldots 6$. The fermion bilinears transform as the adjoint $\textbf{66}$ of $SO(12)$. 
Similarly, $C_{12}$ is charged under the conjugate spinorial $\overline{\textbf{32}}$. Together with the 
three $SU(2)_{k=1}$ currents $J^3,J^\pm$, which are $SO(12)$-singlets, the above currents form the adjoint representation $\textbf{133}$ of $E_7$.

As before, it is convenient to display the spectrum and the spectral flow explicitly in a concrete example. 
To this end, we consider the $\mathcal{N}_4=2$ model  (\ref{N=2_SelfDual}) with enhanced $E_7\times SU(2)\times E_8$ gauge
 symmetry, presented in Section \ref{GeneralWL}. This model arises via a Gepner map from an $\mathcal{N}_4=4$ Type II
 compactification on $T^2\times T^4/\mathbb{Z}_2$, in which 2 spacetime supersymmetries arise from each of the left- and 
right-moving sectors. It corresponds precisely to an $N=(4,4)\oplus(2,2)$ compactification. In terms of covariant conformal blocks, the Gepner map is realized as:
\begin{align}
\frac{1}{2}\sum\limits_{a,b}{(-)^{a+b+ab}\theta[^a_b]^2\theta[^{a+h}_{b+g}]\theta[^{a-h}_{b-g}]}~\rightarrow~\frac{1}{2}\sum\limits_{k,\ell}{\theta[^k_{\ell}]^6\theta[^{k+h}_{\ell+g}]\theta[^{k-h}_{\ell-g}]}.
\end{align}

In the free-field description, the $\hat{c}=4$ system realizing the $N_L=4$ SCFT, is built out of 2 complex 
fermions $\Psi^{7,8}$ which are bosonized as $e^{\pm iH^j(z)}$, with $j=7,8$. The spectral-flow currents are then constructed out of the free fields as:
\begin{align}\label{SF_N=4}
		C_{12}(z)e^{\pm \frac{i}{2}( H^7(z)-H^8(z) )}~\propto~e^{\pm i\frac{1}{\sqrt{2}}H(z)},
\end{align}
where we emphasize above that, by the properties of $N_L=4$ SCFT, the spectral-flow current has to carry $\pm\frac{1}{2}$ units of $J^3$-charge.

As before, we perform the linear field redefinition:
\begin{align}\nonumber
		&Y(z)= H^6(z)\\ \nonumber
		&X(z)= (H^7(z)+H^8(z))/\sqrt{2}\\
		&H(z)= (H^7(z)-H^8(z))/\sqrt{2}~,
\end{align}
where the scalars $Y,X,H$ are still free and, in particular, $i\partial Y(z)$ is identified with
 the $U(1)$-charge of the $\hat{c}=2$ system, while $J(z)=\frac{i}{\sqrt{2}}\partial H(z)$ is the Cartan charge of the $SU(2)_{k=1}$ current algebra of the $\hat{c}=4$, $N_L=4$ system.

We are now in the position to consider the action of the spectral-flow in the untwisted scalar spectrum. The starting point is the vertex operator in the vectorial representation of $SO(12)$:
\begin{align}\label{v12v4}
		\Psi^A(z)e^{\pm i H^j(z)}e^{-\bar\phi(\bar{z})\pm i\bar{H}_{k}(\bar{z})}~\propto~e^{\pm i \epsilon_j\frac{1}{\sqrt{2}} H(z)},
\end{align}
where $j=7,8$ and $k=3,4$. The $\epsilon_j$ are defined as $\epsilon_7=1$ and $\epsilon_8=-1$.
 We now shift the $SU(2)_{k=1}$ charge $Q$ by $\pm\frac{1}{2}$ units in order to make it 
vanish\footnote{Note that, for an $SU(2)_k$ affine algebra, only the integrable representations $|Q|\leq k/2$ are unitary.}. For concreteness, take $j=7$:
\begin{align}
		e^{\pm iH^7(z)} ~\rightarrow~ e^{\pm\frac{i}{2}(H^7(z)+H^8(z))} .
\end{align}
The flow then takes the $V_{12}V_4$ representation (\ref{v12v4}) into the $S_{12}S_4$ representation in the R-sector of the $N_L=4$ SCFT, with vertex operator:
\begin{align}
		S_{12}(z)e^{\pm \frac{i}{2}(H^7(z)+H^8(z))}e^{-\bar\phi(\bar{z})\pm i\bar{H}_k(\bar{z})} .
\end{align}
Again, the transformation can be verified straightforwardly by considering the action of the spectral-flow current (\ref{SF_N=4}) on the vertex operator (\ref{v12v4}).

We now focus our attention to the twisted fermionic massless spectrum, which is relevant for the 
Spinor-Vector duality map. As in the $\mathcal{N}_4=1$ case, we start from the states transforming in the vectorial representation of $SO(12)$. The vertex operator is:
\begin{align}\label{v12c4}
		\Psi^A e^{\pm\frac{i}{2}\left(H^7(z)-H^8(z)\right)}\Gamma^{1}_{(+)}(z,\bar{z})e^{-\frac{1}{2}\bar\phi(\bar{z})+\frac{i}{2}\bar{H}_0(\bar{z})\pm\frac{i}{2}\left(\bar{H}_1(\bar{z})+\bar{H}_2(\bar{z})\right)},
\end{align}
where again $A=1,\ldots,6$.
It involves the invariant twist-field contribution $\Gamma^{h=1}_{(+)}(z,\bar{z})$, which 
starts with conformal weight $(\frac{1}{4},\frac{1}{4})$ and is associated to the topological contribution of the $h=1$-twisted $\Gamma_{(4,4)}$-lattice with definite (positive) $\mathbb{Z}_2$-parity:
\begin{align}\label{twisted44lattice}
		\Gamma^{h=1}_{(s)} = 
\frac{1}{2^2 \eta^4\bar\eta^4}\sum\limits_{g=0,1}\sum\limits_{\gamma,\delta=0,1}{(-)^{\left(\frac{1-s}{2}\right)g}\,\theta[^\gamma_\delta]^2\theta[^{\gamma+1}_{\delta+g}]^2\times\bar\theta[^\gamma_\delta]^2\bar\theta[^{\gamma+1}_{\delta+g}]^2}.
\end{align}

Under the spectral flow, the $V_{12}C_4 O_{16}$ representation (\ref{v12c4}) will
 be mapped into the spinorial $S_{12}O_4 O_{16}$. This can be seen by shifting the $SU(2)_{k=1}$ charge by $\delta Q=\mp\frac{1}{4}$ units so that it vanishes:
\begin{align}
		e^{\pm \frac{i}{2}(H^7(z)-H^8(z))} \rightarrow \textbf{1}(z),
\end{align}
where by the identity operator $\textbf{1}(z)$, we imply not only the vacuum
 representation but also its higher excitations (with even 2d fermion parity). Together, they build up the fermionic $O_4$-representation and one recovers the vertex operator in the spinorial representation of $SO(12)$:
\begin{align}\label{s12o4}
		S_{12}(z)\textbf{1}(z)\Gamma^{1}_{(+)}(z,\bar{z}) e^{-\frac{1}{2}\bar\phi(\bar{z})+\frac{i}{2}\bar{H}_0(\bar{z})\pm\frac{i}{2}(\bar{H}_1(\bar{z})+\bar{H}_2(\bar{z}))} .
\end{align}

We will now carry out the analysis explicitly by constructing the spectral-flow 
currents in the twisted sector and applying them on the vertex operators of the states.
 As before, the special $\mathbb{Z}_2$-nature of the twist permits us to avoid the twist-field 
formalism and represent the relevant $\Gamma^{h=1}_{(\pm)}(z,\bar{z})$ contributions entirely via 
free fermion characters. Performing the summation and projections in (\ref{twisted44lattice}), we find the explicit form of this representation of the twisted vertex operators :
\begin{align}\nonumber
		\Gamma^{h=1}_{(+)}(z,\bar{z})~=~& \left\{~ O_4 S_4 \bar{O}_4\bar{S}_4 ~\oplus~ O_4 C_4 \bar{O}_4\bar{C}_4 ~\oplus~ V_4 S_4 \bar{V}_4\bar{S}_4 ~\oplus~ V_4 C_4 \bar{V}_4\bar{C}_4 \phantom{\frac{1}{2}}\right.\\ \label{twistFplus44}
		& \left. \phantom{\frac{1}{2}} \oplus~ S_4 O_4 \bar{S}_4\bar{O}_4 ~\oplus~ S_4 V_4 \bar{S}_4\bar{V}_4 ~\oplus~ C_4 O_4 \bar{C}_4\bar{O}_4 ~\oplus~ C_4 V_4 \bar{C}_4\bar{V}_4 ~\right\}.
\end{align}
\begin{align}\nonumber
		\Gamma^{h=1}_{(-)}(z,\bar{z})~=~& \left\{~O_4 S_4 \bar{V}_4\bar{C}_4 ~\oplus~ O_4 C_4 \bar{V}_4\bar{S}_4 ~\oplus~ V_4 S_4 \bar{O}_4\bar{C}_4 ~\oplus~ V_4 C_4 \bar{O}_4\bar{S}_4 \phantom{\frac{1}{2}}\right. \\ \label{twistFminus44}
		&\left. \phantom{\frac{1}{2}} \oplus~ S_4 O_4 \bar{C}_4\bar{V}_4 ~\oplus~ S_4 V_4 \bar{C}_4\bar{O}_4 ~\oplus~ C_4 O_4 \bar{S}_4\bar{V}_4 ~\oplus~ C_4 V_4 \bar{S}_4\bar{O}_4 ~\right\}.
\end{align}
Out of these, only the twisted `ground states' with conformal weight $(*,\frac{1}{4})$ will be considered in the fusion rules, since only they can give rise to massless states.

We are now ready to construct the spectral-flow currents in the twisted sector.
 As argued in the previous section, the untwisted spectral-flow operators (\ref{SF_N=4}) become 
extended by a \emph{chiral} dressing $\Omega_{(\pm)}^I(z)$, with $I=1,2$, of conformal weight $\Delta_{1,(\pm)}=(\frac{1}{2},0)$, $\Delta_{2,(\pm)}=(1\mp 1,0)$, acting  on the twist-field contribution $\Gamma^{h=1}_{(\pm)}(z,\bar{z})$. The operator $\Omega_{(\pm)}^I(z)$ transforms 
as $\Omega_{(\pm)}^I\rightarrow \pm \Omega_{(\pm)}^I$ under the $\mathbb{Z}_2$-orbifold. Its action on the twisted ground-state vertex operators relevant for the massless spectrum follows the fusion rule:
\begin{align}\nonumber
		&\Omega_{(\pm)}^1(z)~\cdot~\Gamma^1_{(r)}(w,\bar{w}) = \frac{\Gamma^1_{(+)}(w,\bar{w})}{(z-w)^{\frac{1}{2}+\frac{1-r}{4}}}+\frac{\Gamma^1_{(-)}(w,\bar{w})}{(z-w)^{\frac{1-r}{4}}}+\ldots\\ \nonumber
		&\Omega_{(+)}^2(z)~\cdot~\Gamma^1_{(r)}(w,\bar{w}) = \Gamma^1_{(r)}(w,\bar{w})+\ldots\\
		&\Omega_{(-)}^2(z)~\cdot~\Gamma^1_{(r)}(w,\bar{w}) = \frac{\Gamma^1_{(-r)}(w,\bar{w})}{(z-w)^{1-\frac{r}{2}}}+\ldots
\end{align}
Again, in the interest of notational simplicity, we suppress the representation indices and the 
Dirac matrices of the transformation. The ellipsis denotes, as usual, less singular terms. In terms of free bosonic fields, $\Omega_{(\pm)}^I(z)$ can be represented as:
\begin{align}\nonumber
	&\Omega_{(\alpha)}^1(z)= e^{\pm \frac{i}{2}(\Phi_1 -\alpha \Phi_2)\pm \frac{i}{2}(\Phi_3 -\alpha \Phi_4)},\\ \nonumber
	&\Omega_{(+)}^2(z) = \textbf{1}(z),\\ \label{dressing2}
	&\Omega_{(-)}^2(z) = e^{\pm i\left(\frac{1+r}{2}\right)\Phi_1\pm i\left(\frac{1-r}{2}\right)\Phi_2\pm i\left(\frac{1+s}{2}\right)\Phi_3\pm i\left(\frac{1-s}{2}\right)\Phi_4}
\end{align}
where the $\pm$-signs are arbitrary and independent, $\alpha=\pm 1$ is the $\mathbb{Z}_2$-parity of the operator and $r,s=\pm 1$. 
There are two invariant spectral-flow operators responsible for the mapping between the various representations. They are given as the zero mode of the invariant current:
\begin{align}\nonumber
		Q_{\textrm{s.f.}}^I = \oint{\frac{dz}{2\pi i}\left[~z^{1-I/2}(C_{12}C_4)(z)\Omega_{(+)}^I(z)~+~z^{I/2}(S_{12}S_4)(z)\Omega_{(-)}^I(z)~\right]}, \\ 
\end{align}
or explicitly:
\begin{align}	\nonumber
		& Q_{\textrm{s.f.}}^1 = \oint{\frac{dz}{2\pi i}\, z^{1/2}\left( C_{12}C_4 C_4 C_4 + S_{12}S_4 S_4 S_4 \right)(z)} ,\\ \label{MSDScurrent}
		& Q_{\textrm{s.f.}}^2 = \oint{\frac{dz}{2\pi i}\, \left( C_{12}C_4 \textbf{1}_4 \textbf{1}_4 + z\, S_{12}S_4 V_4 V_4 \right)(z)} .
\end{align}
where in the above we made use of the explicit spin-field representations (\ref{dressing2}).

It is now straightforward to consider the action of the spectral-flow charge on the massless $S_{12}O_4O_{16}$-spinorial representation:
\begin{align}
		Q_{\textrm{s.f.}} \cdot \left[S_{12}O_4 ~\Gamma^{1}_{(+)}e^{-\frac{1}{2}\bar\phi+\frac{i}{2}\bar{H}_0}\bar{S}_4\bar{O}_4\right]|0\rangle  
		= \left[\,V_{12}C_4 \Gamma^{1}_{(+)}\,+\, O_{12}S_4 \Gamma^{1}_{(-)} \,\right] e^{-\frac{1}{2}\bar\phi+\frac{i}{2}\bar{H}_0}\bar{S}_4\bar{O}_4 |0\rangle .
\end{align}
We, thus, see that the spinorial representation of $SO(12)$ is precisely mapped into the 
vectorial $V_{12}C_4 O_{16}$, together with the accompanying singlet $O_{12}S_4O_{16}$. The exact map between the 
representations is, again, seen to arise from the spectral-flow of the twisted $N_L=4$ SCFT. The matching of the numbers of massless degrees of freedom is, hence, a byproduct of the spectral-flow map:
\begin{align}
		\begin{array}{l| l l}
				\textrm{Spinorial}~ & ~S_{12}O_4  & \rightarrow~~ 2^{6-1}\times (4\times 4) \\ \hline \hline
				\textrm{Vectorial}~ & ~V_{12}C_4  & \rightarrow~~ 12\times 2\times(4\times 4)\\ \hline
				\textrm{Singlet}~ & ~O_{12}S_4  & \rightarrow~~ 1\times 2\times (8\times 2\times 4)\\
		\end{array}
\end{align}

As was the case in the previous section, here as well the spectral-flow is responsible for a number of 
supersymmetric-like identities, realized internally in the left-moving sector. In the present case, however, 
the analogous identities are not those of `conventional' supersymmetry, but rather exhibit the precise degeneracy 
structure of MSDS constructions \cite{MSDS}, \cite{ReducedMSDS}. The reason for this is that the relevant orbifold block:
\begin{align}
		Z[^{\,1\,}_{\,g\,}]~=~\frac{1}{2^2\eta^{12}\bar\eta^{4}}\left[\sum\limits_{\ell=0,1}{(-)^{\ell} \theta[^k_\ell]^6 \theta[^{k+1}_{\ell+g}]^2}\right] \left[\sum\limits_{\gamma,\delta=0,1}{\theta[^\gamma_\delta]^2\theta[^{\gamma+1}_{\delta+g}]^2\bar\theta[^{\gamma}_{\delta}]^2\bar\theta[^{\gamma+1}_{\delta+g}]^2}\right]
\end{align}
corresponds to boundary conditions for the free fields that satisfy the conditions \cite{ReducedMSDS} for the MSDS spectral-flow 
operator to have a well-defined action on the spectrum. Indeed, the operator (\ref{MSDScurrent}) is exactly the 
invariant $\mathbb{Z}_2\times \mathbb{Z}_2\times\mathbb{Z}_2$-truncation of the maximal MSDS charge, whose vertex operator
 is proportional to the spin-field $C_{24}(z)$ of $SO(24)$. The spectral-flow is, hence, identified with the spectral-flow 
of constructions with MSDS structure, the only difference being that the `equal' characters between `vectors' and `spinors' are, again, summed rather than subtracted.

Let us exhibit these identities in a systematic way. Let us pick the contribution form the $Z^{h=1}_{(+)}$-orbifold
 block of positive $\mathbb{Z}_2$-parity, which couples to the $P_0$ twisted right-moving characters and which is the only $h=1$ sector which contains massless fermions:
\begin{align}\label{z_plus_h=1}
		Z^{h=1}_{(+)}(\epsilon)= \left(O_{12}S_4 + \epsilon\, C_{12}V_{4}\right) \Gamma^{h=1}_{(-)} + \left(V_{12}C_4+ \epsilon\, S_{12}O_4\right)\Gamma^{h=1}_{(+)}.
\end{align}
Here we explicitly kept the dependence on $\epsilon=\pm 1$, which distinguishes the cases where 
local worldsheet supersymmetry is present ($\epsilon=-1$) or absent ($\epsilon=+1$). As was argued 
in the previous section, the two cases are characterized by the fact that only from the sector with
 local worldsheet supersymmetry can spacetime fermions arise. In our case, where $N_L=4$ is embedded 
inside the `bosonic' sector of the Heterotic string, there are no cancellations between the vectorial
 and spinorial contributions to the partition function. However, it will be instructive to explicitly
 display the identities in the $\epsilon=-1$ case, which illustrate the spectral-flow of the representations
 in a particularly clear way and, at the same time, may give rise to considerable algebraic simplifications, as will be shown below.

The direct evidence of an MSDS spectral-flow at work can be obtained straightforwardly by calculating the $\epsilon=-1$ 
contributions that couple to the various right-moving lattice pieces $\bar{V}_4\bar{C}_4$, $\bar{V}_4\bar{S}_4$, $\ldots \in \Gamma^{h=1}_{(+)}$ :
\begin{align}\nonumber
	\bar{V}_4\bar{C}_4\cdot &\left[ ~\left(O_{12}S_4- C_{12}V_4\right) O_4 S_4 + \left(V_{12}C_4- S_{12}O_4\right) V_4 C_4 \phantom{\frac{}{}} ~\right] ~=~ 4\, \bar{V}_4\bar{C}_4 ~,\\ \nonumber
	\bar{V}_4\bar{S}_4\cdot &\left[ ~\left(O_{12}S_4- C_{12}V_4\right) O_4 C_4 + \left(V_{12}C_4- S_{12}O_4\right) V_4 S_4 \phantom{\frac{}{}} ~\right] ~=~ 4\, \bar{V}_4\bar{S}_4 ~,\\ \nonumber
	\bar{O}_4\bar{C}_4\cdot &\left[ ~\left(O_{12}S_4- C_{12}V_4\right) V_4 S_4 + \left(V_{12}C_4- S_{12}O_4\right) O_4 C_4 \phantom{\frac{}{}} ~\right] ~=~ 0 ~,\\ \nonumber
	\bar{O}_4\bar{S}_4\cdot &\left[ ~\left(O_{12}S_4- C_{12}V_4\right) V_4 C_4 + \left(V_{12}C_4- S_{12}O_4\right) O_4 S_4 \phantom{\frac{}{}} ~\right] ~=~ 0 ~,\\ \nonumber
	\bar{C}_4\bar{V}_4\cdot &\left[ ~\left(O_{12}S_4- C_{12}V_4\right) S_4 O_4 + \left(V_{12}C_4- S_{12}O_4\right) C_4 V_4 \phantom{\frac{}{}} ~\right] ~=~ 4\, \bar{C}_4\bar{V}_4 ~,\\ \nonumber
	\bar{S}_4\bar{V}_4\cdot &\left[ ~\left(O_{12}S_4- C_{12}V_4\right) C_4 O_4 + \left(V_{12}C_4- S_{12}O_4\right) S_4 V_4 \phantom{\frac{}{}} ~\right] ~=~ 4\, \bar{S}_4\bar{V}_4 ~,\\ \nonumber
	\bar{S}_4\bar{O}_4\cdot &\left[ ~\left(O_{12}S_4- C_{12}V_4\right) C_4 O_4 + \left(V_{12}C_4- S_{12}O_4\right) S_4 O_4 \phantom{\frac{}{}} ~\right] ~=~ 0 ~,\\ 
	\bar{C}_4\bar{O}_4\cdot &\left[ ~\left(O_{12}S_4- C_{12}V_4\right) S_4 V_4 + \left(V_{12}C_4- S_{12}O_4\right) C_4 O_4 \phantom{\frac{}{}} ~\right] ~=~ 0 ~.
\end{align}
Of course, if we restrict our attention to the massless spectrum, only the weight-$(*,\frac{1}{4})$ contributions 
in $Z^{h=1}_{(+)}$ are relevant and the corresponding identities are the `supersymmetric-like' ones, with a vanishing r.h.s. as shown above. It is, nevertheless, possible to utilize the above identities in the form:
\begin{align}
		V_{12}C_4 \Gamma^{h=1}_{(+)} + O_{12}S_4 \Gamma^{h=1}_{(-)} ~=~ S_{12}O_4 \Gamma^{h=1}_{(+)} + C_{12}V_4 \Gamma^{h=1}_{(-)} + 4\, \overline{\mathcal{Z}}(\bar{q}),
\end{align}
which permits algebraic simplifications in the partition function. Here, the right-moving 
contribution $\overline{\mathcal{Z}}(\bar{q})$ corresponds to effectively spurious modes and is eliminated by imposing level matching, or by integration of the Teichm\"{u}ller parameter.

This concludes our analysis of the twisted $N_L=4$ case. It becomes clear how the duality between $SO(12)$ spinors and vectors is a direct consequence of the 
spectral-flow of the $N_L=4$ SCFT. A number of chiral identities between current algebra characters illustrate this spectral-flow and can, in some cases, be utilized to algebraically simplify the partition function.


\section{Narain lattices \label{NarainLattices}}

Partition functions offer a very compact way to encode the spectrum of 
a conformal field theory or string compactification. Some of the phenomena
discussed in the previous sections become more intuitive if we use a  
more explicit description provided by the relation between states
and the Narain lattice \cite{Narain}. For
toroidal compactifications the Narain lattice encodes the possible winding,
momentum and gauge charges as a function of the background 
fields \cite{NarainSarmadiWitten}. While
twists which can be realized as pure shifts of the Narain lattice just relate
one toroidal compactification to another, more general twists lead to 
orbifold models
with reduced supersymmetry and reduced gauge groups. As reviewed in Section
\ref{SVreview} the orbifold partition function depends on 
lattices obtained from the original Narain lattices by shifting and 
projection onto invariant states. However, as long as all gauge symmetries
come from the untwisted sector, the twist invariant part of the
Narain lattice contains the full information about symmetry
enhancement and symmetry breaking. This allows to discuss the 
continuous interpolation between models in a very explict way.

\subsection{Review of Narain lattices}

Modular invariance implies that the Narain lattice $\Gamma=\Gamma_{(22,6)}$ 
underlying a toroidal compactification of the heterotic string to four
dimensions must be an even self-dual lattice
with respect to the quadratic form of type $(+)^{22}(-)^6$. 
All such lattices form a single continuous family and can be deformed
into one another by $SO(22,6)$ transformations. String vacua related
by $SO(22)\times SO(6)$ transformations are equivalent, so that the
moduli space locally takes the form
\[
\frac{SO(22,6)}{SO(22) \times SO(6)} \;.
\]
The following standard basis of the Narain lattice
provides an explicit parametrization of this moduli space in terms of the
background fields $G_{IJ}, B_{IJ}, A_I=(A_I^a)$,  $I=1,\ldots, 6$,
$a=1,\ldots, 16$  \cite{Ginsparg}:
\begin{eqnarray}
\overline{k}^I &=& \left( 0 , \frac{1}{2} e^{*I} ;\frac{1}{2} e^{*I} 
\right) \;,\nonumber \\
k_I &=& \left( A_I , e_I + B_{IJ} e^{*J} - (\frac{1}{4} A_I \cdot A_J)
e^{*J} ;- e_I + B_{IJ} e^{*J} - (\frac{1}{4} A_I \cdot A_J)
e^{*J}\right) \;, \nonumber \\
l_a &=& \left( \alpha_a,  - (\alpha_a \cdot A_K) \frac{1}{2} e^{*K} ;
- (\alpha_a\cdot  A_K) \frac{1}{2} e^{*K} \right) \;.\label{NarainBasis}
\end{eqnarray}
Here $\{ e_I \}$ is a basis of the compactification lattice $\Lambda$,
$\{ e^{*I} \}$ is the dual basis of the dual lattice $\Lambda^*$, and 
$\alpha_a$, $a=1, \ldots, 16$ are a set of simple roots of $E_8 \times E_8$.
The basis vectors have the following mutual scalar products:
\[
\bar{k}^I \cdot k_J = \delta^I_J \;,\;\;\;
l_a \cdot l_b = C_{ab} \;,
\]
where $C_{ab}$ is the Cartan matrix of $E_8 \times E_8$, and where
all scalar products which are not displayed are zero. The 
background fields parametrizing the moduli space are the 
lattice metric $G_{IJ} = e_I \cdot e_J$ of $\Lambda$, the
antisymmetric tensor field $B_{IJ}$ and the Wilson lines
$A_I=(A_I^a) \in \mathbb{R}^{16}$.
We will take the lattice $\Lambda$ to be generic throughout, 
and for simplicity we will 
only consider models with vanishing $B$-field, $B_{IJ}=0$.
The integer expansion coefficients $n_I$, $m^I$, $Q^a$ of a 
Narain vector $v\in \Gamma_{(22,6)}$ with respect to this basis
\[
v = n_I \bar{k}^I + m^I {k}_I + Q^a l_a \;,
\]
are the momentum, winding and gauge quantum numbers of the 
corresponding states.

If we switch off the Wilson lines, $A_I=0$, the lattice basis 
takes the simple form
\begin{eqnarray}
\bar{k}^I 
&=& \left( 0 , \frac{1}{2} e^{*I} ; \frac{1}{2} e^{*I} \right) \;, \nonumber \\
{k}_I &=& \left( 0 , e_I  ; - e_I \right) \;, \nonumber \\
l_a &=& \left( \alpha_a,  0_6 ;0_6 \right) \;.
\label{NarainBasisNoWL}
\end{eqnarray}
On the special locus $A_I=0$ the Narain lattice factorizes,
$\Gamma=\Gamma_{16} \Gamma_{(6,6)}$ and the generic
gauge symmetry $U(1)^{28}$ is enhanced to $E_8\times E_8 \times U(1)^{12}$.
Non-abelian gauge symmetries are identified by looking for Narain vectors
of the purely left-moving form $(p_L;0_6)$, with $p_L^2=2$. 
Such vectors automatically form the root system of a semi-simple 
ADE-type Lie algebra. 

Orbifold twists are often defined with respect to a particular subspace
of the Narain moduli space. One of the standard constructions, which 
we use for the $\mathbb{Z}_2$ orbifold, is to combine
an automorphism (rotation or reflection)  
$\theta_{(6)}$ of the compactification lattice 
$\Lambda$ with a shift $\delta_{(16)}$ in the 
$E_8\times E_8$ root lattice. Modular invariance imposes constraints 
on the allowed pairs $(\theta_{(6)}, \delta_{(16)})$. The induced action on 
the Narain lattice is obvious as long as Wilson lines are switched 
off, $A_I=0$, so that the lattice takes the factorized form 
$\Gamma=\Gamma_{16} \Gamma_{(6,6)}$. In particular, the `gauge twist'
$\delta_{(16)}$ acts on $\Gamma$ by the trivially extended shift vector
\begin{equation}
\label{ShiftVNarain}
\delta = (\delta_{(16)}, 0_6 ;0_6 ) \;.
\end{equation}
It is clear that this vector is no longer an admissible shift vector
for deformed Narain lattices obtained by switching on Wilson lines. 
The reason is that a shift vector of order $N$ must have the property
that $N\delta \in \Gamma$, while $k\delta \not\in \Gamma$ for $0<k<N$. 
To check whether any given vector is in $\Gamma$, we only need to check
whether its scalar product with the basis vectors is integer, because
$\Gamma$ is self-dual. Assuming that $N\delta$, with $\delta$ given by 
(\ref{ShiftVNarain}) is in the lattice generated by (\ref{NarainBasisNoWL})
it is clear that it cannot be in the lattice generated by the
deformed basis (\ref{NarainBasis}), except possibly for special values
of the $A_i$. However, it is easy to see that the gauge twist $\delta_{(16)}$
is consistent with the most general continuous Wilson lines. We just
have to modify the extended shift vector (\ref{ShiftVNarain}) by
applying to it the same $SO(22,6)$ boost which relates the two 
bases (\ref{NarainBasisNoWL}) and (\ref{NarainBasis}):
\begin{equation}
\label{DefShiftV}
\delta = \left( \delta_{(16)} , -(\delta \cdot A_K) \frac{1}{2}e^{*K};
-(\delta \cdot A_K) \frac{1}{2} e^{*K} \right) \;.
\end{equation}
To check that (\ref{DefShiftV}) is an admissible shift vectors for 
all values $A_I$ of the Wilson lines, we note that since $\delta_{(16)}$
is an admissible shift of $\Gamma_{16}$, 
\[
N \delta_{(16)} = \sum_{a=1}^{16} Q^a \alpha_a \;,
\]
where $Q^a \in \mathbb{Z}$, and where the simple roots $\alpha_a$ of 
$E_8 \times E_8$ form a lattice basis of $\Gamma_{16}$.  For $A_I=0$, 
it follows that 
\[
N\delta = (\delta_{(16)} , 0_6;0_6) = \sum_{a=1}^{16} Q^a l_a \;.
\]
If we switch on the most general Wilson lines, the 
basis vectors are deformed according to (\ref{NarainBasis}), 
and the resulting deformed shift vector is indeed (\ref{DefShiftV}).
Thus a gauge twist acting as a pure shift does not restrict the
allowed values of the Wilson lines, which in particular remain continuous.

The $\mathbb{Z}_2$ orbifold considered in Section \ref{SVreview} combines
an order 2 automorphism $\theta_{(6)}$ 
of the compactification lattice $\Lambda$, with an
order 2 shift of the gauge lattice $\Gamma_{16}$, which corresponds
to the standard embedding (of the spin into the gauge connection). 
Since $\theta_{(6)}$ acts as identity on two directions and as a reflection
on the other four, the underlying torus must factorize as $T^2\times T^4$.
The action of the twist on the gauge lattices is by a pure shift,
and therefore we can switch on Wilson lines along the $T^2$, which can have
arbitrary continuous values, as discussed above.
In other words the  $\mathbb{Z}_2$ twist acts 
consistently on all Narain lattices of the form
\[
\Gamma = \Gamma_{(16+2,2)} \Gamma_{(4,4)} \;,
\]
where $\Gamma_{(16+2,2)}$ combines the $T^2$ and gauge degrees of freedom,
whereas $\Gamma_{(4,4)}$ is the momentum/winding lattice of the $T^4$.
It is manifest that the (untwisted) moduli space of this family of
orbifold models is given by (\ref{UntwistedModuli}). This demonstrates
how the same conclusion can be reached using either the partition function or
the Narain lattice, and illustrates how a simpler, more geometric 
description arises by using the Narain lattice.

\subsection{Gauge symmetries and continuous Wilson lines}

Another advantage of the Narain lattice is that it is very easy to 
trace patterns of symmetry enhancement and symmetry breaking. For toroidal
models one has complete control of the possible non-abelian gauge symmetries,
and for orbifold models the same is true at least for the untwisted sector
which is obtained from the underlying toroidal model by projection onto
invariant states. While in general the contributions of twisted sectors
need to be investigated explicitly, one can verify, using the partition 
function, for the orbifolds considered in this article that no
gauge symmetry enhancements can arise from the twisted sectors. Therefore
we focus on the projected, untwisted sector and on the underlying toroidal
model in the following.

\begin{figure}[t]
\unitlength1cm
\begin{picture}(9,3)
\put(1,1){\usebox{\Eachterw}}
\end{picture}
\caption{The extended Dynkin diagram of $E_8$. \label{E8}}
\end{figure}
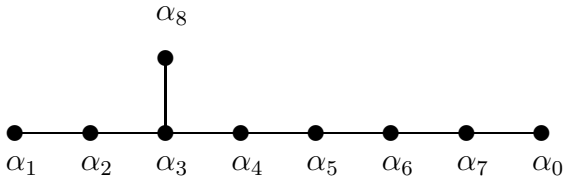

Let us first review how the breaking of $E_8\times E_8$ by 
Wilson lines can be controlled and parametrized, following
\cite{TM1}.  For simplicity
we only consider one $E_8$ factor, with simple roots $\alpha_a$, 
$a=1\ldots, 8$. There is a well known algorithm for constructing 
successively the maximal (regular) subalgebras of a simple Lie algebra,
which works by successively removing dots from the extended Dynkin
diagram \cite{Cahn}. The extended Dynkin diagram of $E_8$ is obtained (as for 
any simple Lie algebra) by adding the 
lowest root
\[
\alpha_0 = - \sum_{a=1}^8 k^a \alpha_a \;,
\]
to the Dynkin diagram. The coefficients 
$(k^a)=(2,4,6,5,4,3,2,3)$, together with $k^0=1$ are known
as the Kac labels. The extended Dynkin diagram of $E_8$ is displayed
in Figure \ref{E8}. 
For concreteness, we specify an explicit choice of simple roots
for $E_8$ together with the resulting lowest root:
\begin{eqnarray}
\alpha_1 &=& \left( - \frac{1}{2}\;,- \frac{1}{2}\;,- \frac{1}{2}\;,
- \frac{1}{2}\;,- \frac{1}{2}\;,- \frac{1}{2}\;,- \frac{1}{2}\;,
- \frac{1}{2}\right)  \;,\nonumber \\
\alpha_2 &=& \left( 0,0,0,0,0,1,0,1\right) \;, \nonumber \\
\alpha_3 &=& \left( 0,0,0,0,1,-1,0,0 \right) \;, \nonumber \\
\alpha_4 &=& \left( 0,0,0,0,-1,0,1,0 \right) \;, \nonumber \\
\alpha_5 &=& \left( 0,0,0,1,0,0,-1,0 \right) \;, \nonumber \\
\alpha_6 &=& \left( 0,0,1,-1,0,0,0,0 \right) \;, \nonumber \\
\alpha_7 &=& \left( 0,1,-1,0,0,0,0,0 \right) \;, \nonumber \\
\alpha_8 &=& \left( 0,0,0,0,0,1,0,-1 \right) \;, \nonumber \\
\alpha_0 &=& \left( 1,-1,0,0,0,0,0,0 \right) \;. \label{E8basis}
\end{eqnarray}
The breaking of any of the two $E_8$ groups in Narain models can be
controlled by monitoring which of the Narain vectors corresponding to 
simple roots are `projected out.' Since 
\[
( A_I, *_6 ; *_6 ) \cdot (\alpha_a , 0_6;0_6) = A_I \cdot \alpha_a \;,
\]
it is clear that a massless boson with charges corresponding to 
$\alpha_a$ is present in the spectrum if and only if
$A_I \cdot \alpha_a \in \mathbb{Z}$. If this is condition is 
violated by changing
the Wilson lines continuously, then this state acquires a mass,
controlled by the Wilson lines, through the Higgs mechanism.

As an example we consider the breaking of $E_8$ to $E_7 \times U(1)$ 
and $E_7\times SU(2)$. By inspection of Figure \ref{E8}, the removal of the
dot corresponding to $\alpha_7$ results in the Dynkin diagram of
$E_7 \times SU(2)$, while removing in addition the dot corresponding
to  $\alpha_0$ gives the Dynkin diagram of $E_7$. `Removing dots'
can be implemented by chosing Wilson lines which have non-integer
scalar products with the dots one wants to remove but integer 
scalar products with the dots one wants to keep \cite{TM1}. 
For concreteness, consider
switching on a Wilson line of the form $A_1 = \lambda \alpha_7^*$,
where $\lambda$ is a continuous parameter, and where $\alpha_7^*$ 
is the dual of the seventh root (= the seventh fundamental weight),
\[
\alpha_a^* \cdot \alpha_b = \delta_{ab} \;.
\]
In the representation we have chosen
\[
\alpha_7^* = (-1,1,0,0,0,0,0,0) \;.
\]
Depending on the value of $\lambda$, there are the following three cases:
\begin{enumerate}
\item
If $\lambda \in \mathbb{Z}$, 
then the Wilson line $A_1$ has integer scalar product
with all simple roots of $E_8$, and the gauge symmetry is not reduced. 
For $\lambda \not= 0$ this
corresponds to a T-duality transformation, where the Narain lattice is
mapped to itself.\footnote{The T-duality group $SO(22,6,\mathbb{Z})$ 
consists
precisely of those isometries that act automorphically on the lattice,
and thus can be `undone' by a basis transformation.}
\item
If $\lambda = \frac{1}{2} \mbox{  mod  } \mathbb{Z}$, 
then the Wilson line
has an integer scalar product with all simple $E_8$ roots, except
with $\alpha_7$, and the scalar product with the lowest root $\alpha_0$
is integer. The resulting root system corresponds to $E_7\times SU(2)$,
which shows that these discrete Wilson lines break $E_8$ to this
maximal subgroup.
\item
If $\lambda$ is not an integer multiple of $\frac{1}{2}$, then the
Wilson line has integer scalar product with all simple roots of
$E_8$ except $\alpha_7$, and the scalar product with the lowest root
$\alpha_0$
is not integer. Since the number of Cartan generators is not reduced, 
the unbroken subgroup is $E_7 \times U(1)$.
\end{enumerate}
Thus the continuous Wilson line $A_1 = \lambda \alpha_7^*$ generically
breaks $E_8$ to $E_7 \times U(1)$, but at special values this is
re-enhanced to the maximal subgroup $E_7 \times SU(2)$. Moreover 
$\lambda \simeq \lambda + 1$ by T-duality. 

Discrete Wilson lines, where a multiple of the Wilson line lies
in $\Gamma_{16}$, can be re-interpreted as orbifolds acting by
pure shifts. For the case at hand, note
that the discrete Wilson line
\[
A_1 = - \frac{1}{2} \alpha_7^* = \left( \frac{1}{2}, - \frac{1}{2}, 
0,0,0,0,0,0\right) \;,
\]
which breaks $E_8$ to $E_7 \times SU(2)$, can be re-interpreted as
a shift vector. Indeed, $\delta_{(16)}=-\frac{1}{2} \alpha_7^*$
is an admissible shift vector of order 2, because $2\delta_{(16)} 
\in \Gamma_{16}$. Note that $\alpha_7^* \in \Gamma_{16}$, because
the $E_8$ root lattice is selfdual. This is an explicit example
where a pure shift can be re-interpreted as a discrete change
of background fields. 
Incidentially, the shift $\delta_{(16)}=-\frac{1}{2} \alpha_7^*$
is the bosonic version of
the standard embedding gauge twist (\ref{StdEmb})
of the $\mathbb{Z}_2$ orbifold. Note, however, that if an orbifold
acts by a combination of an automorophism of $\Lambda$ 
with a gauge shift, it
need not be true any more that the 
gauge shift can be replaced by a discrete background field. 
In particular,
once the shift $\delta_{(16)}=-\frac{1}{2} \alpha_7^*$
becomes part of the definition of the $\mathbb{Z}_2$
orbifold it cannot be replaced by a discrete Wilson line any more. 
However, the $\mathbb{Z}_2'$ orbifold can be re-interpreted in terms
of background fields, because it acts by a pure shift.

Let us next investigate the symmetry breaking patterns of the
$\mathbb{Z}_2$ orbifold model. As far as gauge symmetry breaking
is concerned, the shift 
\begin{equation}
\label{gauge_shift}
\delta_{(16)}=-\frac{1}{2} \alpha_7^* = 
\left( \frac{1}{2}, - \frac{1}{2}, 
0,0,0,0,0,0,0,\ldots, 0\right)
\end{equation}
has the same effect as the Wilson line discussed above and breaks
$E_8 \times E_8$ to $E_7 \times SU(2) \times E_8$. This is clear
because at the $E_8 \times E_8$ point the invariant states of the
untwisted 
sector correspond to Narain lattice vectors which have integer
scalar products with $(\delta_{(16)}, 0, \ldots, 0)$. We will now
investigate what happens to this sector if we switch on 
a Wilson line, which we parametrize as
\[
A_1 = (a_1, a_2| b_1, b_2, b_3, b_4, b_5, b_6| c_1, c_2, \ldots c_8) \;.
\]
The symmetry breaking of the second $E_8$ by the $c_1,\ldots, c_8$
follows the same pattern as in toroidal models. Therefore we will
focus on the first $E_8$ and frequently suppress 
the vector components corresponding
to the second $E_8$. The explicit basis (\ref{E8basis}) is not
adapted to the subgroup $E_7 \times SU(2)$ unbroken by the twist, 
but instead to the maximal subgroup $SO(16)$, since all roots
are chosen to be either adjoint weights of $SO(16)$, 
($\alpha_0, \alpha_2, \ldots, \alpha_8$) or conjugate spinor weigths
($\alpha_1$). It is convenient to further decompose
the $E_8$ roots with respect to the subgroup
\[
SO(4) \times SO(12) \subset SO(16) \subset E_8 \;.
\]
As we will see explicitly below, using $SO(4) \simeq SU(2) \times
SU(2)$,  this allows us to fit states into representations of $E_7\times
SU(2)$ through the chain of subgroups
\[
SU(2) \times SU(2) \times SO(12) \subset SU(2) \times E_7 \subset E_8\;.
\]
In terms of $SO(16)$ weights, the 240 roots of $E_8$ 
are obtained by combining the 112 roots of $SO(16)$
\[
(0 \cdots \pm 1 \cdots \pm 1 \cdots 0)
\]
(where precisely two entries are non-vanishing and take values $\pm 1$)
and the 128 conjugate spinor weights
\[
(\underbrace{\pm \frac{1}{2} , \pm \frac{1}{2} , \cdots , 
\pm \frac{1}{2} }_{\mbox{even}}) \;.
\]
(where the number of $(-)$-signs is even). In the untwisted sector
of the orbifold, all vectors which do not have integer scalar
products with the shift vector (\ref{gauge_shift})
are projected out. Out of the 240 $E_8$ root vectors the following 
128 vectors survive the projection,
\begin{eqnarray}
(\pm 1, \pm 1 | 0,0,0,0,0,0) \;,&& \nonumber \\
(0,0,| \cdots \pm 1 \cdots \pm 1 \cdots) \;,&& \nonumber \\
 \left(\underbrace{\pm \frac{1}{2}, \pm \frac{1}{2}}_{\rm even}| 
\underbrace{\pm \frac{1}{2}, 
\pm \frac{1}{2}, \pm \frac{1}{2}, \pm \frac{1}{2}, 
\pm \frac{1}{2}, 
\pm \frac{1}{2}}_{\rm even}\right) \;, &&
\end{eqnarray}
and it is easy to see that they span the $(0,0)$ and $(c,c)$ conjugacy
classes of $SO(4) \times SO(12)$, where $(0)$ denotes the class of the
adjoint and $(c)$ denotes the class of the conjugate spinor representation. 
Taking into account that none of the $E_8$ Cartan generators is projected
out, we have precisely the right number of states to fill the 
adjoint representation $(3,1) \oplus (1,133)$ of $SU(2) \times E_7$. 
The weight vectors of this representation can be seen explicitly when
performing a rotation by 45 degree in the first two entries, which 
corresponds to the isomorphism 
$SO(4) \rightarrow SU(2) \times SU(2)$: 
\[
\begin{array}{|l|l|l|} \hline
\mbox{Weights} & SU(2) \times SU(2) \times  SO(12) & SU(2) \times E_7 \\ \hline
\hline
(\pm \sqrt{2}, 0,0,0,0,0,0,0) & (3,1,1) &  (3,1) \\ \hline 
(0, \pm \sqrt{2},0,0,0,0,0,0) & (1,3,1) & \\
(0,0, \cdots \pm 1 \cdots \pm 1 ) &(1,1,66)  & (1,133) \\ 
(0, \pm \frac{1}{2}\sqrt{2}, 
\underbrace{\pm \frac{1}{2},\pm \frac{1}{2},\pm \frac{1}{2},
\pm \frac{1}{2},\pm \frac{1}{2},\pm \frac{1}{2}}_{\rm even}) & (1,2,32)
 & \\
\hline
\end{array}
\]
With this completely explicit description of the gauge group at the
point of maximal symmetry, we can now easily  study the effect of continuous
Wilson lines.  We parametrize the Wilson line in the rotated basis as
\[
\tilde{A}_1 = (\tilde{a}_1, \tilde{a}_2, b_1, b_2, b_3, b_4, b_5, b_6) \;,
\]
where
\[
\tilde{a}_1 = \frac{1}{2} \sqrt{2} (a_1 - a_2) \;,\;\;\;
\tilde{a}_2 = \frac{1}{2} \sqrt{2} (a_1 + a_2) \;.
\]
Now we use that a vector corresponds to a 
massless gauge boson  in the untwisted sector of the orbifold
if and only if it has an integer scalar product with the Wilson 
line.\footnote{This was developed systematically in \cite{TM2}.}
The following cases can occur.
\begin{enumerate} 
\item
If $\sqrt{2} \tilde{a}_1$ is integer, then the first $SU(2)$ factor of
$SU(2) \times SU(2) \times SO(12)$ is unbroken. Otherwise
it is broken.
\item
If $\sqrt{2} \tilde{a}_2$ is integer, then the second $SU(2)$ factor of
$SU(2) \times SU(2) \times SO(12)$ is unbroken. Otherwise
it is broken. If the second $SU(2)$ is broken
there cannot be an unbroken $E_7$.
\item
If $b=(b_1, b_2, b_3, b_4, b_5, b_6)$ is a weight of $SO(12)$, 
then the $SO(12)$ subgroup of $SU(2) \times SU(2) \times SO(12)$
is unbroken. Otherwise it is broken and then in particular the 
$E_7$ symmetry is broken. 
\item
If $b=(b_1, b_2, b_3, b_4, b_5, b_6)$ is a weight in the
adjoint conjugacy class of $SO(12)$, and if $\sqrt{2} \tilde{a}_2$
is an integer, then the $E_7$ is unbroken. Note that if we only require
$b$ to be a weight of $SO(12)$, this only implies 
an unbroken $SU(2) \times SO(12)$ subgroup.  But
if $b$ is an adjoint weight, then 
the additional weights required 
to extend the root system to from $SU(2)\times SO(12)$ to  $E_7$ 
are present, because scalar products between adjoint weights
and weights in any conjugacy class are integer valued
for ADE-type Lie algebras.
\item
We remark that by tuning the Wilson line, the $SO(12)$ factor can be broken
to any regular subgroup, in the way described in \cite{TM1,TM2}.
\end{enumerate}

As an explicit example, the one-parameter family
of Wilson lines
\[
A_1 (\lambda) = (\lambda, 0 | 1-\lambda, 0, \cdots, 0|1,0,\ldots,
0) \;,\;\;\;
0\leq \lambda \leq 1\;,
\]
interpolating between the two discrete Wilson lines
(\ref{WL+}) and (\ref{WL-}), which correspond to the $\mathbb{Z}_2 \times
\mathbb{Z}_2'$ orbifold with discrete torsion $\epsilon = +1$ and
$\epsilon = -1$, respectively. Note that we now include
the second $E_8$. Comparing the last 8 entries of the Wilson line
to (\ref{E8basis}) and Figure \ref{E8}, we see that the second $E_8$ is broken to $SO(16)$. 
The gauge shift already breaks the first $E_8$ to $SU(2) \times E_7$,
and to investigate the further breaking by the Wilson line, we rewrite
it in the $E_7$-adapted basis
\[
\tilde{A}_1 (\lambda) = \left(\frac{\sqrt{2}}{2} \lambda , \frac{\sqrt{2}}{2} 
\lambda  | 1-\lambda, 0, \cdots, 0|1,0,\ldots,
0 \right) \;,\;\;\;
0\leq \lambda \leq 1\;.
\]
Using the above analysis, we see that for $0<\lambda<1$ the
subgroup $SU(2)\times SU(2)$ is broken to $U(1) \times U(1)$, while
$SO(12)$ is broken to $U(1)\times SO(10)$. 
For $\lambda =0,1$ the group $SU(2) \times
SU(2) \times SO(12)$ is unbroken, but the additional weights needed
for enhancement to $SU(2) \times E_7$ are projected out. Thus
we have a family of models with gauge group
\[
U(1)^3 \times SO(10) \times SO(16) \;,\;\;\;0<\delta<1 \;,
\]
which is enhanced to 
\[
SU(2) \times SU(2) \times SO(12) \times SO(16) \;\;\;
\mbox{for}\;\;\;\delta =0,1 \;,
\]
which are the two models related by spinor-vector duality.

\section{Heterotic $K3\times T^2$ compactifications \label{K3section}}

It is well known that the singularities of the orbifold $T^4/\mathbb{Z}_2$
can be deformed to obtain a smooth K3 surface \cite{AspinwallK3}. 
To put our results
into perspective, we will now review some results about 
K3 compactifications of the heterotic string \cite{KachruVafa,MorVaf,LSTY}.
Anomaly cancellation requires that non-trivial gauge fields 
are switched on along the K3. The precise condition is that the
Euler number 24 of K3 is cancelled by the instanton number of
the gauge field configuration. 
Geometrically, this corresponds
to the choice of an $E_8\times E_8$ vector bundle $V$ over K3,
with second Chern class $c_2(V)=24$.  Since the gauge group
has two simple factors, the gauge bundle is a sum $V_1\oplus V_2$,
and one is free to distribute the total instanton number between
the two bundles, 
\[
c_2(V_1) + c_2(V_2) = 24 = \chi_{K3} \;.
\]
The resulting family of models is parametrized by an integer $k$,
\[
(c_2(V_1), c_2(V_2)) = (12+k, 12-k) \;\;\;\;k=0,1,2,\ldots 12\;.
\]
By heterotic-type IIA duality, these models are equivalent 
to compactifications of type-IIA string theory on a family
of Calabi-Yau three-folds which are elliptic fibrations over the
Hirzebruch surfaces $F_k$. Moreover the members of the family are
related by going through loci of enhanced symmetry, which on the
type-II side correspond to singularities of the $F_k$ basis. 

One example is the heterotic K3 compactification
with standard embedding. Here the spin connection on K3 is
identified with an $SU(2)$ subgroup of one of the $E_8$, and the
unbroken gauge group is the commutant $E_7 \times E_8$ of this subgroup.
Together with the abelian factors from further reduction on $T^2$ this
results in a gauge group $E_7 \times E_8 \times U(1)^4$.  
Since all instantons are valued in the same $E_8$-factor, this corresponds
to taking $k=12$ above. Taking all 24 instantons to be valued in the
same $SU(2)$ subgroup is a very special choice which leaves the gauge 
group as large as possible. A generic distribution of the instantons
within an $E_8$ factor breaks it completely. 

The gauge group can also be reduced in another way. 
As in any $N=2$ gauge theory, one can `go to the Coulomb branch'
by turning on generic vacuum expectation values for the scalars
in the four-dimensional vector multiplets. This breaks the gauge 
group to the maximal abelian subgroup $U(1)^{7+8+4} = U(1)^{19}$,
and only neutral hypermultiplets remain massless. The scalars
in these hypermultiplets are the moduli of the K3 surface
and of the gauge bundle.
The (quaternionic) dimensions of these moduli spaces are known to 
be 20 and 45, respectively, so that this model has a rank 19 abelian 
gauge group
and 65 hypermultiplets on the Coulomb branch.

In contrast, 
the $\mathbb{Z}_2$ orbifold model
 considered in this article is the compactification 
of the heterotic string on the orbifold limit
$T^4/\mathbb{Z}_2$ of K3 with standard embedding, which was first studied
in \cite{Walton}. At the point of maximal symmetry this
results in a model with 
gauge group $E_7 \times SU(2)\times E_8 \times U(1)^4$ together
with charged and neutral hypermultiplets. For orbifolds going
to the Coulomb branch corresponds to switching on generic
Wilson lines, which breaks the gauge group to $U(1)^{20}$ and
makes all charged hypermultiplets massive. The only remaining
massless states are 
the $\mathcal{N}_4=2$ gravity multiplet, the dilaton vector 
multiplet and
18 vector multiplets and 4 hypermultiplets corresponding to the
untwisted moduli space (\ref{UntwistedModuli}). 
To relate this model to a smooth K3 compactification, 
one must first go to an enhancement locus where at least
one $SU(2)$ factor is present, because a smooth K3 compactification 
requires a non-vanishing instanton number. Switching on instantons
breaks (at least) one $SU(2)$ factor and the resulting gauge 
group has (at most) rank 19. One specific route for going from 
an heterotic orbifold to a smooth heterotic K3 compactification
was described in \cite{KachruVafa}. In the orbifold model one
first goes to the $E_7\times SU(2)$ locus, and then moves to the
Higgs branch. This means to give vacuum expectation values 
to scalars in hypermultiplets. Each such 
hypermultiplet combines with a vector multiplet into a massive, long
(non-BPS) vector multiplet. This mechanism is able to give mass to neutral
vector multiplets, and therefore it reduces the rank of the gauge 
group. In the example described in \cite{KachruVafa}
3 hypermultiplets are used to Higgs the $SU(2)$. After going to the
Coulomb branch they obtain a model with gauge group $U(1)^{19}$
and 65 hypermultiplets, which  is the generic spectrum of the 
heterotic string on a smooth K3 surface with standard embedding.


\section{Conclusions}\label{Conclusions}

In this paper we studied some of the conformal properties of Spinor-Vector duality, aspiring to trace back its CFT origin. In particular, after reviewing the duality map, we demonstrated how (discrete or even continuous) Wilson lines may be turned on to give masses to the vectorial or spinorial representations of the GUT gauge group. This provided a realization of the duality map as arising from different deformations of the same initial `parent' theory (the S-V self-dual point), which corresponds precisely to points of exceptional gauge symmetry enhancement.

The fact that the number of massless degrees of freedom in the theory with massless vectorials (accompanied by the singlets) was found to precisely coincide with the number of massless states in the SV-dual theory with massless spinorials was the most serious indication that there is an underlying spectral-flow at work. The enhancement points are marked by the appearance of global $N_L=2$ and $N_L=4$ superconformal algebras, which can be regarded as embeddings of the $N=2$, $N=4$ SCFTs of Type II theories into the `bosonic' sector of the Heterotic string, via the Gepner map. 

The superconformal properties and spectral-flow of the $N_L=2$ and $N_L=4$ SCFTs, which are relevant for $\mathcal{N}_4=1$ and $\mathcal{N}_4=2$ spacetime supersymmetry, respectively, were analyzed in some detail. In particular, the invariant spectral-flow operator was constructed explicitly in the twisted sectors and its action was shown to give rise to a specific isomorphism between the current algebra representations. A byproduct of this exact map is the presence of a series of identities, which may be employed both to illustrate the flow between the representations, as well as to algebraically simplify the partition function.

The unexpected and non-trivial result has been the identification of the spectral-flow operator in the $\mathcal{N}_4=2$ case, with the MSDS spectral-flow operator, which otherwise arises as a target-space symmetry in very special 2d string constructions. The MSDS constructions have been recently employed as candidate models in order to probe the early non-geometrical hot temperature phase of the universe. Their thermal interpretation and marginal deformations were discussed in \cite{DeformedMSDS}, in relation to the construction of Hagedorn- and tachyon-free theories. Quite recently, their special symmetric structure was utilized in \cite{HybridCosmo}, in order to construct a stringy thermal model whose induced cosmological evolution is simultaneously free of initial gravitational-type singularities as well as Hagedorn-type instabilities of high temperature string theory, while remaining within the perturbative domain at every stage of the evolution.

The appearance of the MSDS structure in the context of Heterotic $N=(4,4)$-compactifications is highly non-trivial and merits a careful study on its own.
What is more, through the Gepner map, the current algebra of conventional supersymmetry is precisely mapped into the MSDS current algebra, manifested as the algebra of the twisted $N_L=4$ spectral-flow operators. The implications of this spectacular result remain to be investigated at greater depth in future work. However, the present observation already sheds some light into the realizations of the MSDS-structure and the nature of its algebra.

\bigskip
\medskip
\leftline{\large\bf Acknowledgements}
\medskip

We are grateful to C.~Angelantonj, N.~Toumbas and especially to C.~Kounnas for illuminating and inspiring discussions. IF and MT would also like to thank K.~Christodoulides for several fruitful discussions and clarifying remarks. AEF would like to thank the University of Oxford and IF would like to thank the University of Liverpool and the CERN Theory Division for hospitality.
This work is supported in part by an STFC rolling grant ST/G00062X/1.


\appendix
\numberwithin{equation}{section}

\section{Partition Function in the specific $\mathcal{N}_4=2$ example}\label{PartitionFunction}

We describe here the procedure for constructing the partition function of the Heterotic orbifold compactification on $T^2\times T^4/\mathbb{Z}_2$, which was defined in Section \ref{ToyModelReview}. The full partition function is the sum of two (disconnected) orbits of the modular group. 
One first starts with the (unprojected) partition function $Z[^{0,0}_{0,0}]$ in the (fully) untwisted sector:
\begin{align}
	Z[^{0,0}_{0,0}] = \left(\bar{V}_8-\bar{S}_8\right)~ \Gamma_{(4,4)}(G_{IJ},B_{IJ})~\Gamma_{(1,1)}(R)~\left(O_{16}+S_{16}\right)\left(O_{16}+S_{16}\right),
\end{align}

One then acts on $Z[^{0,0}_{0,0}]$ by the non-trivial elements $\{\alpha,\beta,\alpha\beta\}$ of the full orbifold group $\mathbb{Z}_2\times\mathbb{Z}_2'$ to obtain: 
$$
	Z[^{0,0}_{1,0}]=\alpha Z[^{0,0}_{0,0}]~~,~~Z[^{0,0}_{0,1}]=\beta Z[^{0,0}_{0,0}]~~,~~Z[^{0,0}_{1,1}]=\alpha\beta Z[^{0,0}_{0,0}].
$$
The modular orbit is then completed by acting on these elements with the modular group generators $S,T$ (see, for example \cite{carlo2002}):
$$
	Z[^{1,0}_{0,0}]=S \alpha Z[^{0,0}_{0,0}]~~,~~Z[^{0,1}_{0,0}]=S\beta Z[^{0,0}_{0,0}]~~,~~Z[^{1,1}_{0,0}]= S\alpha\beta Z[^{0,0}_{0,0}].
$$
$$
Z[^{1,0}_{1,0}]=TS \alpha Z[^{0,0}_{0,0}]~~,~~Z[^{0,1}_{0,1}]=TS\beta Z[^{0,0}_{0,0}]~~,~~Z[^{1,1}_{1,1}]= TS\alpha\beta Z[^{0,0}_{0,0}].
$$

The second orbit can be easily constructed by picking an $S$-transformed element in the first orbit, such as $Z[^{1,0}_{0,0}]$, and acting on it with the $\beta$-group element so that we obtain an element outside the first orbit:
$$
	Z[^{1,0}_{0,1}] = \beta S\alpha Z[^{0,0}_{0,0}].
$$
The remaining elements in the second orbit can be constructed by the repeated action of the $S,T$-generators on $Z[^{1,0}_{0,1}]$:
$$
	Z[^{0,1}_{1,0}] = S Z[^{1,0}_{0,1}] ~~,~~Z[^{1,0}_{1,1}] = T Z[^{1,0}_{0,1}]~~,~~ Z[^{0,1}_{1,1}] = TS Z[^{1,0}_{0,1}],
$$
$$
	Z[^{1,1}_{1,0}] = ST Z[^{1,0}_{0,1}] ~~,~~ Z[^{1,1}_{0,1}] = TST Z[^{1,0}_{0,1}].
$$


After taking into account the $g,g'$-projections, the untwisted sector can be written as:
\begin{align}\nonumber
	Z_{(0,0)} &= Q_o ~\Lambda_{2m,n} \left[~ \Gamma^{h=0}_{(+)}\left( ~V_{12}V_4 O_{16} +  S_{12}S_{4}S_{16}~\right) + \Gamma^{h=0}_{(-)}\left(~O_{12}O_4 O_{16} + C_{12}C_{4}S_{16}~\right) ~\right]\\ \nonumber ~\\ \nonumber 
	&+ Q_o ~\Lambda_{2m+1,n} \left[~ \Gamma^{h=0}_{(+)}\left( ~S_{12}S_4 O_{16} +  V_{12}V_{4}S_{16}~\right) + \Gamma^{h=0}_{(-)}\left(~C_{12}C_4 O_{16} + O_{12}O_{4}S_{16}~\right) ~\right]\\ \nonumber ~\\ \nonumber
	&+ Q_v ~\Lambda_{2m,n} \left[~ \Gamma^{h=0}_{(-)}\left( ~V_{12}V_4 O_{16} +  S_{12}S_{4}S_{16}~\right) + \Gamma^{h=0}_{(+)}\left(~O_{12}O_4 O_{16} + C_{12}C_{4}S_{16}~\right) ~\right]\\ \nonumber ~\\ \nonumber
	&+ Q_v ~\Lambda_{2m+1,n} \left[~ \Gamma^{h=0}_{(-)}\left( ~S_{12}S_4 O_{16} +  V_{12}V_{4}S_{16}~\right) + \Gamma^{h=0}_{(+)}\left(~C_{12}C_4 O_{16} + O_{12}O_{4}S_{16}~\right) ~\right].\\ \nonumber
\end{align}

Similarly, for the sector twisted under the freely acting $\mathbb{Z}_2'$ :
\begin{align}\nonumber
	Z_{(0,1)} &= Q_o ~\Lambda_{2m,n+\frac{1}{2}} \left[~ \Gamma^{h=0}_{(+\epsilon)}\left( ~O_{12}V_4 V_{16} +  C_{12}S_{4}C_{16}~\right) + \Gamma^{h=0}_{(-\epsilon)}\left(~V_{12}O_4 V_{16} + S_{12}C_{4}C_{16}~\right) ~\right]\\ \nonumber ~\\ \nonumber 
	&+ Q_o ~\Lambda_{2m+1,n+\frac{1}{2}} \left[~ \Gamma^{h=0}_{(+\epsilon)}\left( ~C_{12}S_4 V_{16} +  O_{12}V_{4}C_{16}~\right) + \Gamma^{h=0}_{(-\epsilon)}\left(~S_{12}C_4 V_{16} + V_{12}O_{4}C_{16}~\right) ~\right]\\ \nonumber ~\\ \nonumber
		&+ Q_v ~\Lambda_{2m,n+\frac{1}{2}} \left[~ \Gamma^{h=0}_{(-\epsilon)}\left( ~O_{12}V_4 V_{16} +  C_{12}S_{4}C_{16}~\right) + \Gamma^{h=0}_{(+\epsilon)}\left(~V_{12}O_4 V_{16} + S_{12}C_{4}C_{16}~\right) ~\right]\\ \nonumber ~\\ \nonumber
	&+ Q_v ~\Lambda_{2m+1,n+\frac{1}{2}} \left[~ \Gamma^{h=0}_{(-\epsilon)}\left( ~C_{12}S_4 V_{16} +  O_{12}V_{4}C_{16}~\right) + \Gamma^{h=0}_{(+\epsilon)}\left(~S_{12}C_4 V_{16} + V_{12}O_{4}C_{16}~\right) ~\right].\\ \nonumber
\end{align}

For the sector twisted under the non-freely acting $\mathbb{Z}_2$:
\begin{align}\nonumber
	Z_{(1,0)} &= P_o ~\Lambda_{2m+\frac{1-\epsilon}{2},n} \left[~ \Gamma^{h=1}_{(+)}\left( ~V_{12}C_4 O_{16} +  S_{12}O_{4}S_{16}~\right) + \Gamma^{h=1}_{(-)}\left(~O_{12}S_4 O_{16} + C_{12}V_{4}S_{16}~\right) ~\right]\\ \nonumber ~\\ \nonumber 
	&+ P_o ~\Lambda_{2m+\frac{1+\epsilon}{2},n} \left[~ \Gamma^{h=1}_{(+)}\left( ~S_{12}O_4 O_{16} +  V_{12}C_{4}S_{16}~\right) + \Gamma^{h=1}_{(-)}\left(~C_{12}V_4 O_{16} + O_{12}S_{4}S_{16}~\right) ~\right]\\ \nonumber ~\\ \nonumber
		&+ P_v ~\Lambda_{2m+\frac{1-\epsilon}{2},n} \left[~ \Gamma^{h=1}_{(-)}\left( ~V_{12}C_4 O_{16} +  S_{12}O_{4}S_{16}~\right) + \Gamma^{h=1}_{(+)}\left(~O_{12}S_4 O_{16} + C_{12}V_{4}S_{16}~\right) ~\right]\\ \nonumber ~\\ \nonumber 
	&+ P_v ~\Lambda_{2m+\frac{1+\epsilon}{2},n} \left[~ \Gamma^{h=1}_{(-)}\left( ~S_{12}O_4 O_{16} +  V_{12}C_{4}S_{16}~\right) + \Gamma^{h=1}_{(+)}\left(~C_{12}V_4 O_{16} + O_{12}S_{4}S_{16}~\right) ~\right].\\ \nonumber
\end{align}

Finally, the sector twisted under both $\mathbb{Z}_2\times\mathbb{Z}_2'$ is:
\begin{align}\nonumber
	Z_{(1,1)} &= P_o ~\Lambda_{2m+\frac{1-\epsilon}{2},n+\frac{1}{2}} \left[~ \Gamma^{h=1}_{(+\epsilon)}\left( ~O_{12}C_4 V_{16} +  C_{12}O_{4}C_{16}~\right) + \Gamma^{h=1}_{(-\epsilon)}\left(~V_{12}S_4 V_{16} + S_{12}V_{4}C_{16}~\right) ~\right]\\ \nonumber ~\\ \nonumber 
	&+ P_o ~\Lambda_{2m+\frac{1+\epsilon}{2},n+\frac{1}{2}} \left[~ \Gamma^{h=1}_{(+\epsilon)}\left( ~C_{12}O_4 V_{16} +  O_{12}C_{4}C_{16}~\right) + \Gamma^{h=1}_{(-\epsilon)}\left(~S_{12}V_4 V_{16} + V_{12}S_{4}C_{16}~\right) ~\right]\\ \nonumber ~\\ \nonumber
		&+ P_v ~\Lambda_{2m+\frac{1-\epsilon}{2},n+\frac{1}{2}} \left[~ \Gamma^{h=1}_{(-\epsilon)}\left( ~O_{12}C_4 V_{16} +  C_{12}O_{4}C_{16}~\right) + \Gamma^{h=1}_{(+\epsilon)}\left(~V_{12}S_4 V_{16} + S_{12}V_{4}C_{16}~\right) ~\right]\\ \nonumber ~\\ \nonumber 
	&+ P_v ~\Lambda_{2m+\frac{1+\epsilon}{2},n+\frac{1}{2}} \left[~ \Gamma^{h=1}_{(-\epsilon)}\left( ~C_{12}O_4 V_{16} +  O_{12}C_{4}C_{16}~\right) + \Gamma^{h=1}_{(+\epsilon)}\left(~S_{12}V_4 V_{16} + V_{12}S_{4}C_{16}~\right) ~\right].\\ \nonumber
\end{align}

In the above expressions we make use of the decomposition: 
\begin{align}\label{twistLattice_a}
		\Gamma^{h}_{(\pm)} \equiv \frac{1}{2}\left( \Gamma_{(4,4)}[^h_0] \pm \Gamma_{(4,4)}[^h_1] \right),
\end{align}
of the symmetrically twisted $(4,4)$-lattice:
\begin{align}\label{twistLattice_b}
	\Gamma_{(4,4)}[^h_g] = 
\left\{\begin{array}{c l}
	\Gamma_{(4,4)}(G,B) &,~\textrm{for}~(h,g)=(0,0) \\
	\left|\frac{2\eta}{\theta[^{1-h}_{1-g}]}\right|^{4} &,~\textrm{for}~(h,g)\neq(0,0)\\
\end{array}\right.
\end{align}
into linear combinations with a definite $\mathbb{Z}_2$-parity.

\section{Operator Products of $SO(N)$ Spin Fields}\label{OPEs}

The fusion rules in the text can be verified straightforwardly by repeated use of the very useful OPEs involving the spin-fields of $SO(N)$ (see, for example, \cite{Kostelecky}):
\begin{align}
	\psi^a(z) S_{\alpha}(w) &= \frac{\gamma^a_{\alpha\beta}}{\sqrt{2}}\frac{S_{\beta}(w)}{(z-w)^{1/2}}+\ldots\\ \nonumber ~\\ \nonumber
		 S_\alpha(z) S_\beta(w) & = \frac{c_{\alpha\beta}}{(z-w)^{N/8}}+\frac{\gamma^a_{\alpha\beta}}{\sqrt{2}}\frac{\psi^a(w)}{(z-w)^{N/8-1/2}}\\ 	\nonumber	&+\frac{\gamma^{ab}_{\alpha\beta}}{2}\frac{\psi^a\psi^b(w)}{(z-w)^{N/8-1}}+\frac{1}{2\sqrt{2}}\frac{\gamma^{abc}_{\alpha\beta}\,\psi^a\psi^b\psi^c(w)+\gamma^a_{\alpha\beta}\, \partial\psi^a(w)}{(z-w)^{N/8-3/2}}\\
&+\frac{1}{4}\frac{\gamma^{abcd}_{\alpha\beta}\,\psi^a\psi^b\psi^c\psi^d(w)+\gamma^{ab}_{\alpha\beta}\,\partial(\psi^a\psi^b)(w)+\frac{1}{2}c_{\alpha\beta}(\partial\psi^a)\psi^a(w)}{(z-w)^{N/8-2}}+\ldots,
\end{align}
where $c_{\alpha\beta}$ is the charge conjugation matrix in the Dirac representation and the ellipses denote less singular terms.



\bigskip
\medskip

\bibliographystyle{unsrt}

\vfill\eject
\end{document}